\documentclass{PoS}
\ifx\selectfont\undefined%
  \def\mathrm#1{{\rm #1}}
  \def\textrm#1{{\rm #1}}
\fi

\newcommand{\tanbeta}{\ensuremath{\tan\beta}}
\newcommand{\mh}{\ensuremath{\mathrm m_{H}}}
\newcommand{\mA}{\ensuremath{\mathrm m_{A}}}
\newcommand{\hplus}{\ensuremath{\mathrm H^{+}}}
\newcommand{\ifb}{\ensuremath{fb^{-1}}}
\newcommand{\ET}{\ensuremath{E_{T}}}
\newcommand{\pT}{\ensuremath{p_{T}}}
\newcommand{\pz}{\ensuremath{p_{z}}}
\def\GeV{\ifmmode {\mathrm{\ Ge\kern -0.1em V}}\else
                   \textrm{Ge\kern -0.1em V}\fi }%
\newcommand{\SM}{\mbox{Standard Model}}
\newcommand{\bbbar}{\ensuremath{\mathrm {b\overline{\mathrm b}}}}
\newcommand{\ttbar}{\ensuremath{\mathrm {t\overline{\mathrm t}}}}
\newcommand{\csbar}{\ensuremath{\mathrm {c\overline{\mathrm s}}}}
\newcommand{\WWlnln}{\ensuremath{\mathrm {WW \rightarrow \ell \nu \ell \nu}}}
\newcommand{\WWlnqq}{\ensuremath{\mathrm {WW \rightarrow \ell \nu qq}}}
\newcommand{\ZZllll}{\ensuremath{\mathrm {ZZ \rightarrow \ell  \ell \ell \ell}}}
\newcommand{\ZZllqq}{\ensuremath{\mathrm {ZZ \rightarrow \ell  \ell qq}}}
\newcommand{\ZZllnn}{\ensuremath{\mathrm {ZZ \rightarrow \ell  \ell \nu \nu}}}

\title{LHC Higgs Boson searches}

\ShortTitle{LHC Higgs Boson}

\author{\speaker{William Murray}\thanks{On behalf of the ATLAS and CMS collaborations.}\\
        STFC-RAL / CERN\\
        E-mail: \email{bill.murray@stfc.ac.uk}}

\abstract{A summary of the Higgs boson searches by the ATLAS and CMS
  collabrations using 1\ifb\ of LHC data is presented, concentrating
  on the Standard Model Higgs boson. Both experiments
  have the sensitivity to exclude at 95\% CL a 
  \SM\ Higgs boson in most of the Higgs boson mass region between about 130~\GeV\ and
  400~\GeV. The  observed data allow the exclusion of a Higgs Boson of
  mass 
155~\GeV\ to   190~\GeV\ and 295~\GeV\ to 450~\GeV\  (ATLAS) and 
149~\GeV\ to   206~\GeV\ and 300~\GeV\ to 440~\GeV\  (CMS).
The lower limits are not as constraining as might be
  expected due to an excess in both experiments of order 2-3$\sigma$
  which could be related to a low mass Higgs boson or to a
  statistical fluctuation.}

\FullConference{The 2011 Europhysics Conference on High Energy Physics, EPS-HEP 2011,\\
		July 21-27, 2011\\
		Grenoble, Rh\^one-Alpes, France}

\begin{document}

\section{Introduction}

The Higgs boson\cite{Englert:1964et,Higgs:1964ia,Guralnik:1964eu}
search at the LHC has entered its prime. Results in 2010 were
groundbreaking in many areas, especially in the MSSM searches for
neutral Higgs bosons\cite{PhysRevLett.106.231801, Aad:2011rv}, but it is
only with the advent of datasets of the scale of an inverse femtobarn
that the 
possible existence of the Standard Model (SM) Higgs boson can start to be
tested.
 This the experiments have
done with great enthusiasm. 

However, the fast moving nature of the search and the rapidly
improving LHC performance and hence integrated luminosity mean that
this report on these searches is already most interesting as a
historical snapshot of the knowledge at the time of the conference and the
methodologies used by the experiments; the actual results having been
in most cases superseded. This review attempts to present that
snapshot, without updating with later results. The \SM\ Higgs boson
was the star of the conference, and other results are either very
briefly summarised or skipped entirely.

In this document, the word lepton, $\ell$, should normally be interpreted as
referring to electrons or muons and their antiparticles. Limits are
all quoted at 95\% CL.

\section{MSSM Higgs Bosons}
  The Higgs mechanism, the introduction of a complex doublet  field with a
  quartic self-coupling and negative quadratic term, is rather general
  and while the simplest version is employed in the SM, it can be
  extended in many ways. One of the interesting extensions is the 
  addition of a second Higgs doublet, and in particular the so-called `type
  II' doublet\cite{PhysRevD.41.3421} required by supersymmetry.
Within this framework there are 5 physical Higgs scalars, two charged
and three neutral, whose properties are completely defined at tree
level by two parameters, often taken to be \mA\ and \tanbeta ( the
ratio of the vacuum expectation values of the two doublets).
The three neutral bosons are the lighter and heavier scalars,
respectively h and H, and the pseudo-scalar A.

The search for the lightest scalar, h, is in most scenarios closely
related to the SM Higgs Boson search described later, but
this 
section sketches the results of the search for  neutral or charged
MSSM Higgs bosons.

\subsection{Neutral MSSM Higgs Boson searches}
\label{sec:mssmneut}
   The heavy MSSM Higgs bosons, A and H, do not couple to the W and Z bosons,
   but have a coupling to the down-type fermions proportional to
   \tanbeta. Searches at the LHC\cite{epsMSSMatlas} have so far
   focused on the decay mode H/A 
   to $\tau\tau$,  which has a sensitivity roughly proportional to
   \tanbeta$^2$. The production comes either through gluon fusion,
   dominant for low \tanbeta, or from associated production with one
   or more b quarks, which grows proportional to \tanbeta$^2$. The
   CMS collaboration presented results\cite{epsttcms,cms-tt}  (given in
   Fig.~\ref{fig:htautau}), on this search using 
   1.1~\ifb\ of data and a detailed analysis of the production mode.
That is to say, three production modes were considered, gluon fusion,
b-quark associated and vector boson fusion or VBF. The second is
especially appropriate for high \tanbeta\ MSSM Higgs bosons.

  \begin{figure}[h]
\centering
\includegraphics[width=0.7\textwidth]{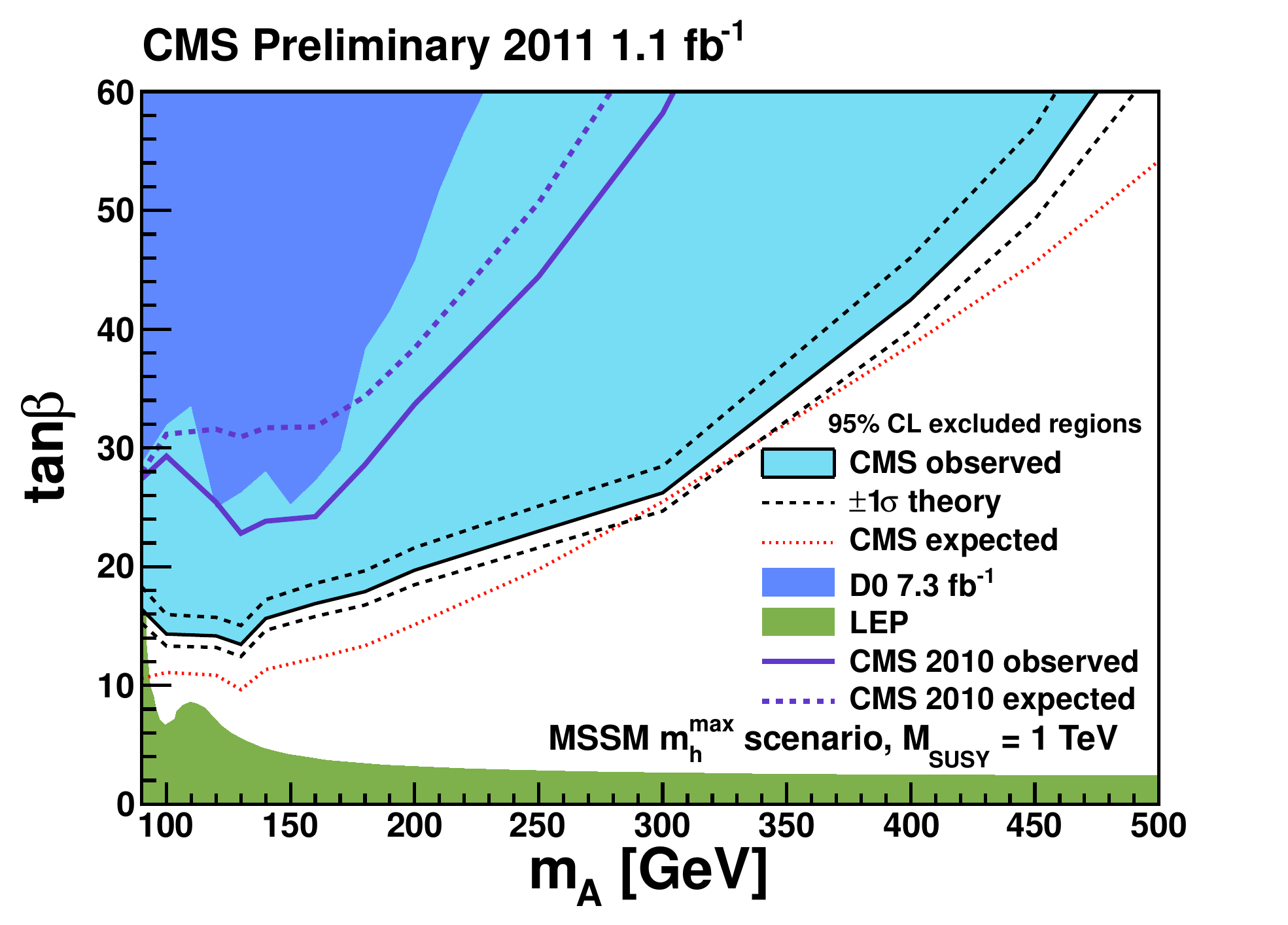}
    \vskip -0.2cm
    \caption[]{Limits on the \mA-\tanbeta\ plane coming from the H/A
      to $\tau\tau$ search in CMS.}
    \label{fig:htautau}
  \end{figure}

The $\tau$ pairs were studied in $e\mu$, $\mu\mu$, $e\tau_{had}$ and
$\mu\tau_{had}$ decay modes, with no evidence seen for Hihhs boson production.

\subsection{Charged MSSM Higgs boson search}
   The charged Higgs boson presents a tempting search target as it is an
   unambiguous indicator of physics beyond the Standard Model. 
The coupling between the charged Higgs and the top quark is strong, but
the phenomenology depends crucially upon the relative masses and
hence which decays into which. If the charged Higgs weighs less than
the top quark it can be produced with a large rate in top decay. If
not the production cross-sections are not yet accessible at the LHC.
  The mass in the MSSM is similar to the heavy neutral Higgs bosons,
  and for many parameters there is greater sensitivity to those.

The production via top decay
   to \hplus$b$ has been searched for, and limits have bee derived on the
   hadronic decay \csbar\cite{atlas-cs} by ATLAS and more
   stringently on the decay 
   $\hplus \rightarrow \tau\nu$\cite{cms-hplus,epshpluscms} by CMS. The
   latter analysis considers three final states: electron plus muon,
   muon plus a hadronic tau, and no leptons plus a hadronic tau. The
   combined limits extracted from the three channels   are shown in
   Fig.~\ref{fig:hplus} in terms of the top decay fraction, which is
   limited to below about 4\% for the masses tested and in the
   \mA-\tanbeta\ plane. 

  \begin{figure}[h]
\centering
\includegraphics[width=0.49\textwidth]{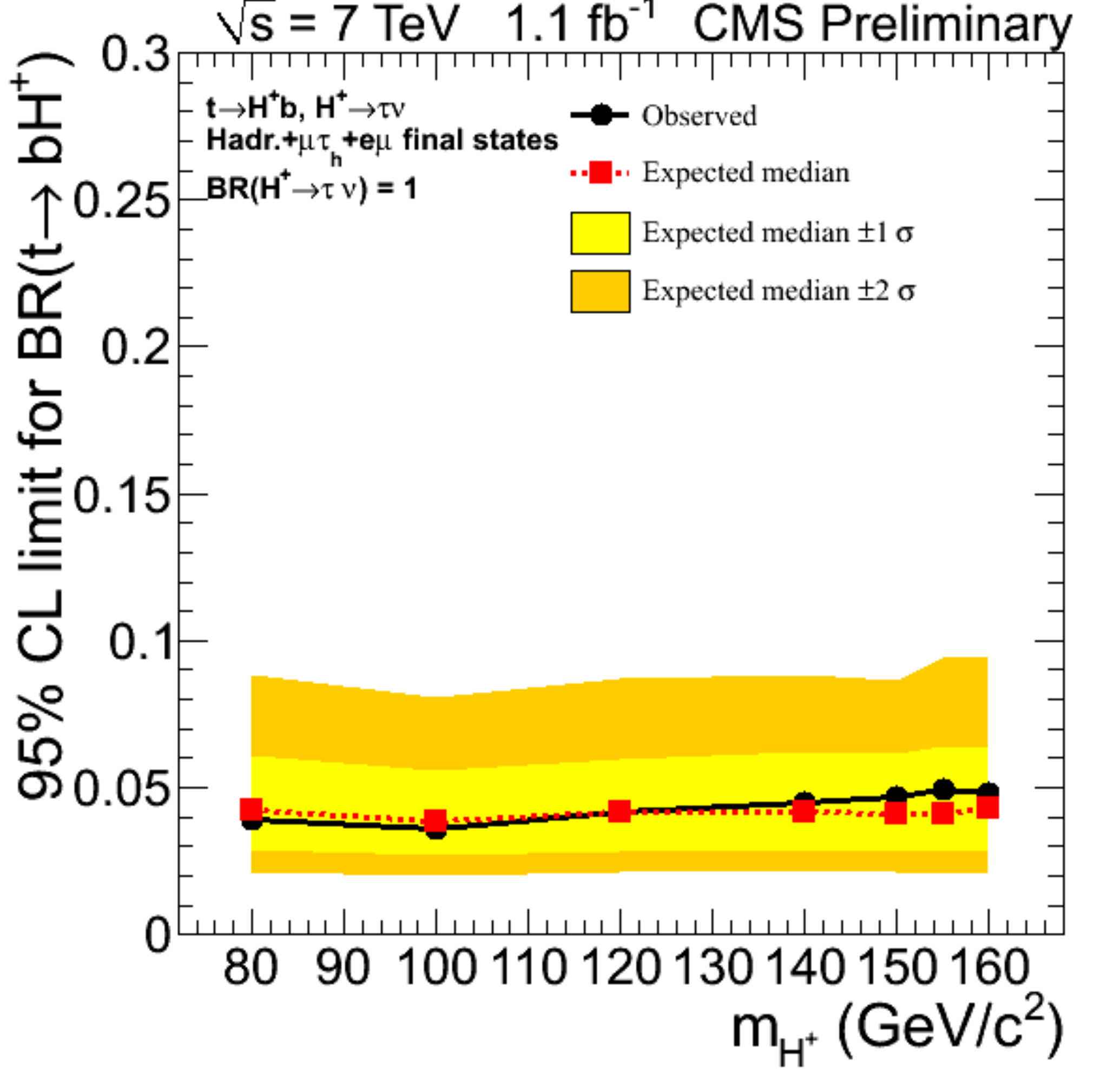}
\includegraphics[width=0.49\textwidth]{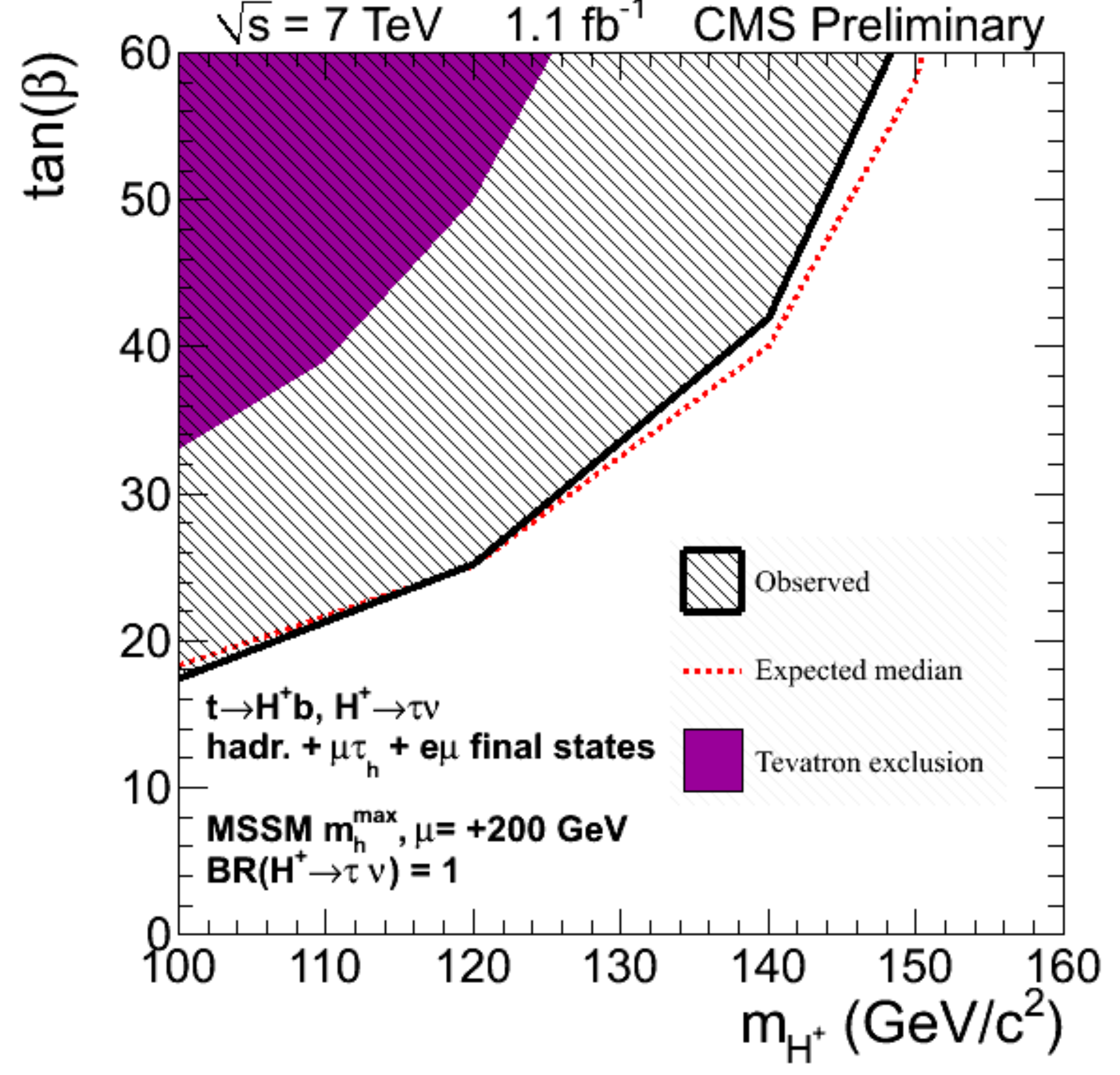}
    \vskip -0.2cm
    \caption[]{Limits on charged Higgs boson production  from CMS, expressed in terms
      of (left) the top branching ratio to charged Higgs boson and (right) the \mA-\tanbeta\ plane.}
    \label{fig:hplus}
  \end{figure}

\section{SM Higgs Boson}
The most promising aspect of the LHC Higgs boson analyses  is the search for the
SM Higgs boson. Radically new sensitivity is available for
this decades-old search, combining information from multiple decay
modes to enhance the overall sensitivity. To roughly summarize,
the individual experiments, with 1\ifb\ each, each have at least one-sigma 
 sensitivity to Higgs bosons with masses between about 120~\GeV\ and
 560~\GeV, and 
 considerably more for most masses. This allows the most sensitive
 test of the Higgs mechanism ever.

Within the \SM, the only unknown aspect of the Higgs boson is its
mass; after that everything is predicted. For Higgs boson masses above about
125~\GeV\ the search is  dominated by the bosonic decay modes,
WW and ZZ. These have so far been searched for in channels where at
least one vector boson decays to leptons: \WWlnln, \WWlnqq,
\ZZllll, \ZZllnn\  and \ZZllqq. For lower masses, down to 110~\GeV,  the
decays to $b\overline{b}$,  
$\tau\tau$ and especially $\gamma\gamma$ play a gradually increasing
role but are not sensitive to the  Higgs boson  at its expected
\SM\ rate with the
data available at this meeting.

The main production modes are, in order of importance to the searches
reported here, gluon fusion, vector boson fusion (VBF) and
associated production with a W or Z boson or a pair of top quarks.
The cross-section calculations have been made by many groups over many
years, but all the numbers used here are collated  in  the LHC Higgs
cross-section working group
report\cite{LHCHiggsCrossSectionWorkingGroup:2011ti}. 

\subsection{Higgs boson decay to $b\overline{b}$}

The dominant decay mode by rate for Higgs bosons masses below 135~GeV
is  to a pair of bottom quarks, but searches 
are complicated by the enormous LHC b-quark production rates in
other processes. The situation is somewhat improved in the associated
production modes, which can also be used to provide a trigger. ATLAS
presented searches in $WH\rightarrow l\nu bb$ and $ZH\rightarrow ll
bb$  modes\cite{epsbbatlas,bbnote}.
The   resulting mass spectra can be seen in Fig~\ref{fig:hbb}.


  \begin{figure}[h]
\centering
\includegraphics[width=0.49\textwidth]{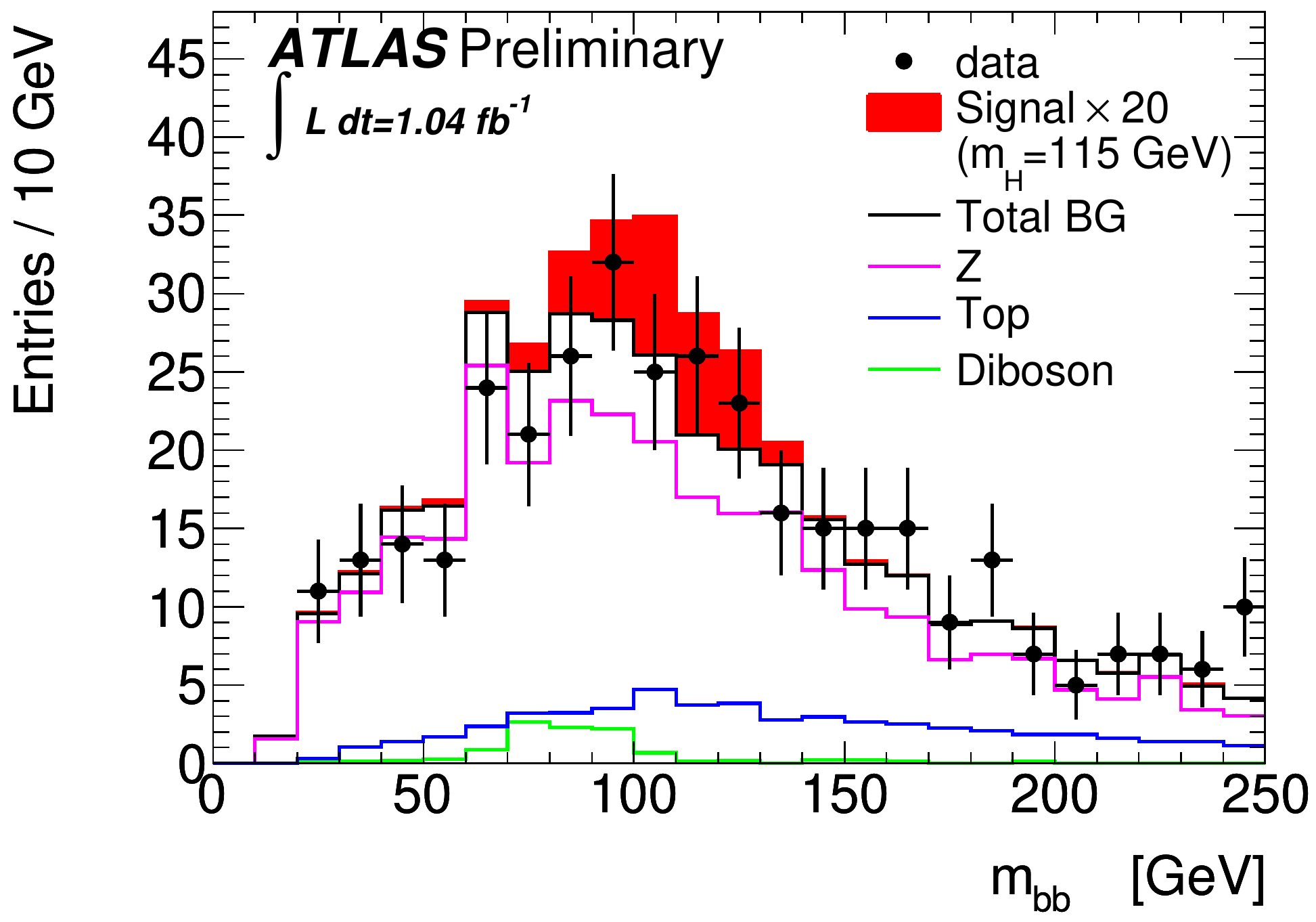}
\includegraphics[width=0.49\textwidth]{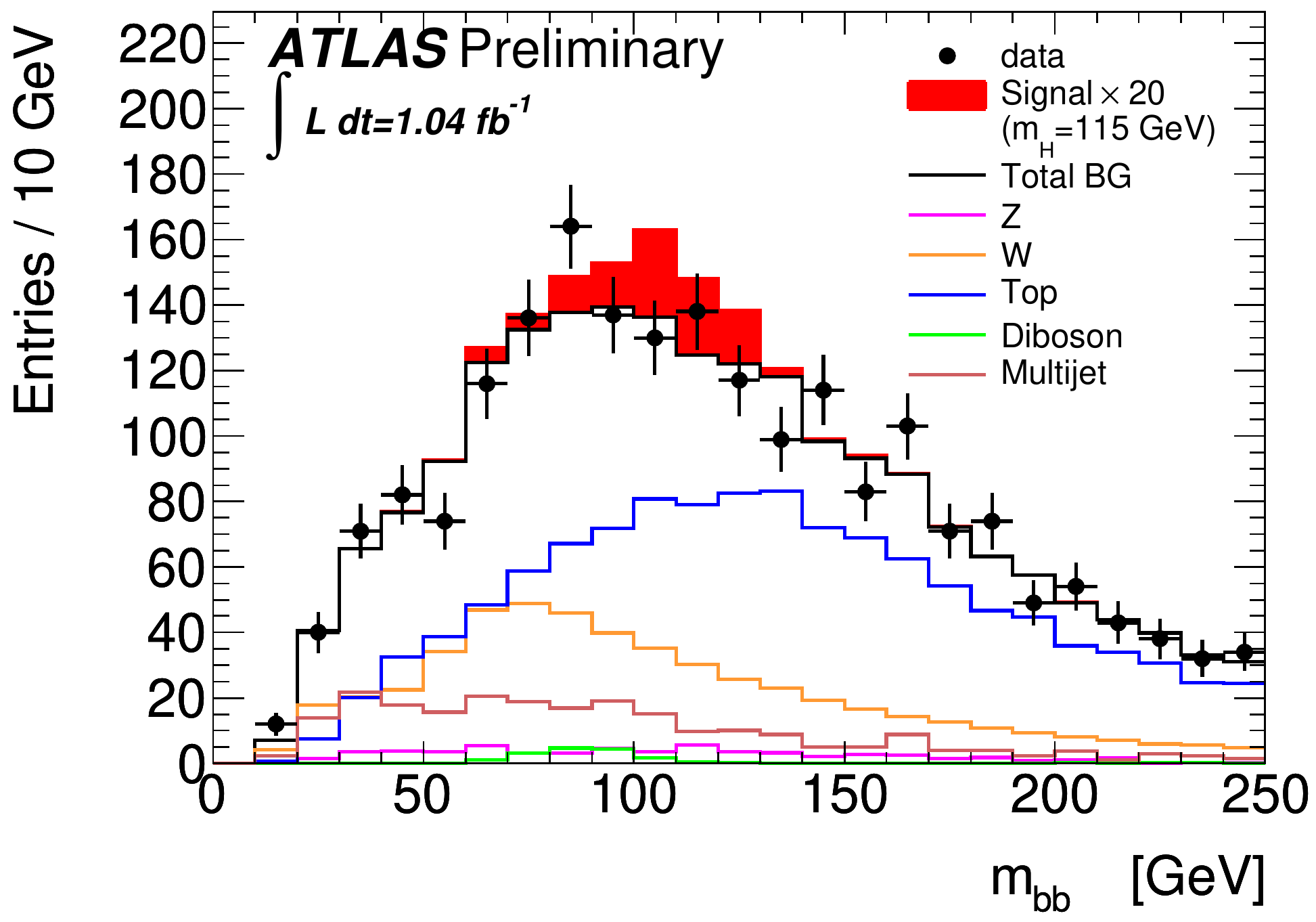}
    \vskip -0.2cm
    \caption[]{Mass spectra found by ATLAS in the ZH (left) and WH
      (right) searches.}
    \label{fig:hbb}
  \end{figure}

These searches would be sensitive to a signal with around fifteen
times the SM rate. Significant improvements will be
needed, and the use of the those events with high \pT\, where the
signal  to background ratio improves, is one route\cite{Butterworth:2008iy}.

\subsection{Higgs boson decay to $\tau\tau$}

The MSSM tau pair searches mentioned in section~\ref{sec:mssmneut} can
also be used to look for a SM Higgs boson. In this case the VBF
production mode  is an important mechanism giving adequate production
rate with sufficient rejection of backgrounds.
ATLAS re-interpreted their MSSM search\cite{epsMSSMatlas} in a SM context.
The CMS collaboration presented results from $e\mu$, $\mu\mu$, $\mu\tau$
and $e \tau$ searches\cite{epsttcms}; the example of the $e\mu$
channel in the VBF production mode is shown in Figure~\ref{fig:htt}.
The analysis 
focussing on  vector boson fusion production is, as expected,
significantly more powerful than the inclusive gluon fusion search,
but 
the use of a central jet veto adds systematic errors to the signal acceptance.

  \begin{figure}[h]
\centering
\includegraphics[width=0.49\textwidth]{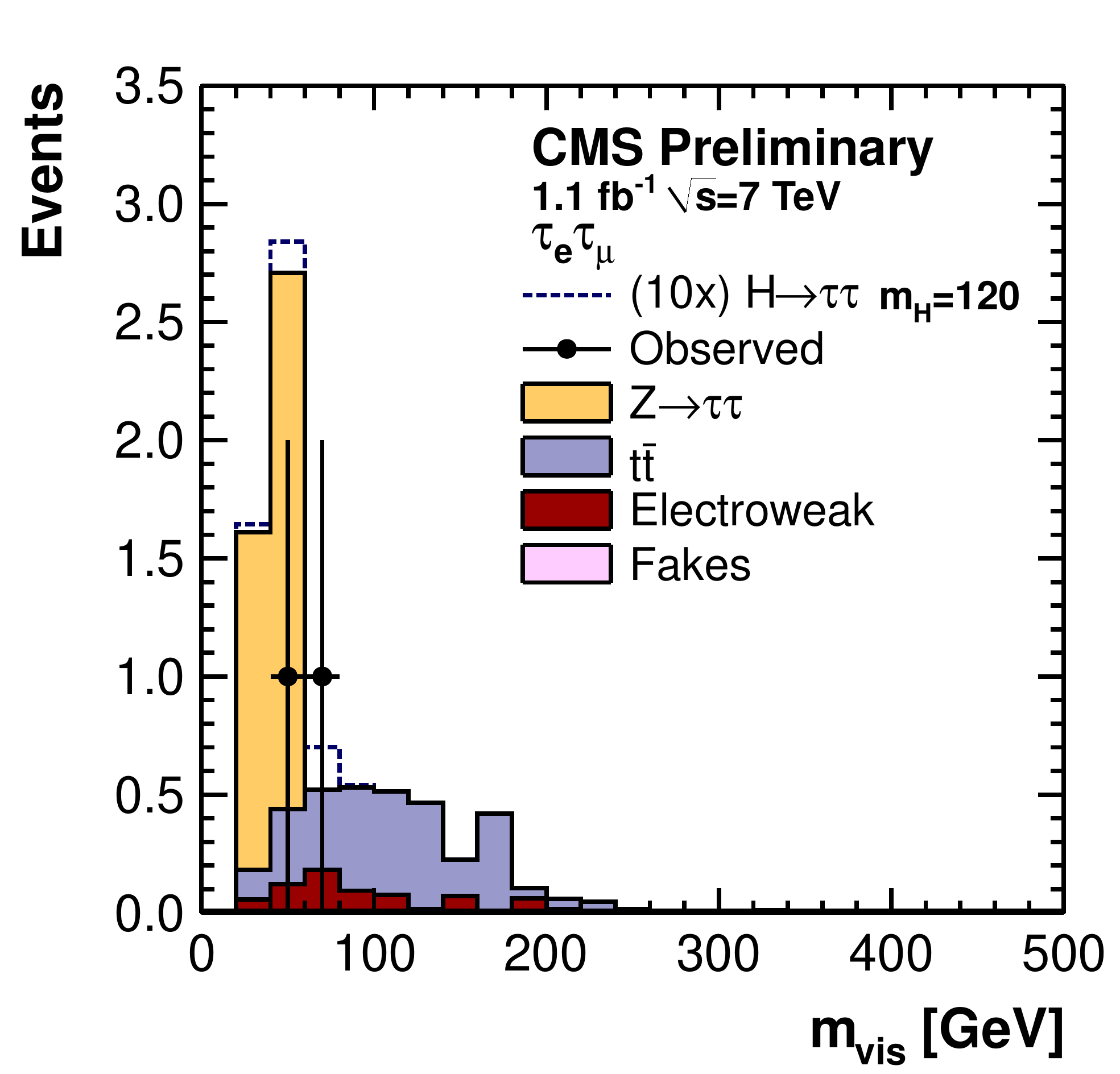}
\includegraphics[width=0.49\textwidth]{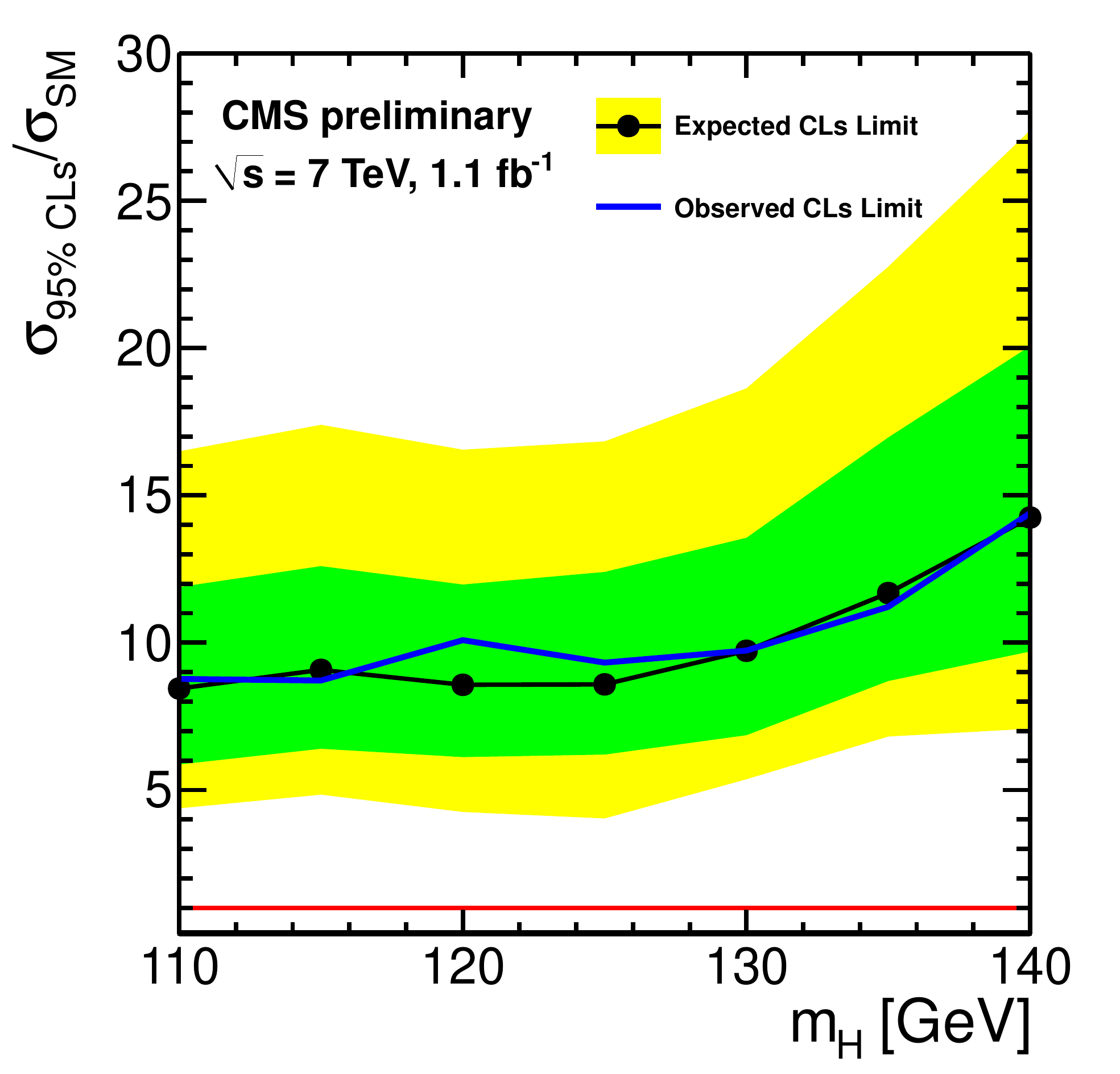}
    \vskip -0.2cm
    \caption[]{Left: The mass spectrum from CMS VBF $H\rightarrow \tau\tau
      \rightarrow e \mu$ search. Right: The limits expected and
      observed in the $H \rightarrow \tau\tau$ search as a function of
      \mh\ in units of the
      \SM\ cross-section. } 
    \label{fig:htt}
  \end{figure}

The CMS search is sensitive to eight or more times the SM Higgs bosons
cross-section, and the observed data match the expectations from background.

\subsection{Higgs boson decay to $\gamma\gamma$}

The requirements of measuring the $\gamma\gamma$ decay
process\cite{epsggcms,epsggatlas,Aad2011452,cms-gamgam} have driven
the 
performance of the   electromagnetic calorimeters of ATLAS and CMS.
The CMS crystal calorimetry offers superior energy resolution (though
complete calibration is still ongoing) while the ATLAS segmented
calorimetry allows a 
simultaneous measurement of the photon angle. This means that the
Higgs candidate  decay vertex location in CMS is identified using
tracking system, while ATLAS relies on the calorimetry.
The  complete mass
spectra of both experiments are shown in Fig~\ref{fig:hgg}, but the
analysis is done in six or eight sub-samples of different signal to
background depending upon the kinematics or quality of the photon candidates.

  \begin{figure}[h]
\centering
\includegraphics[width=0.49\textwidth]{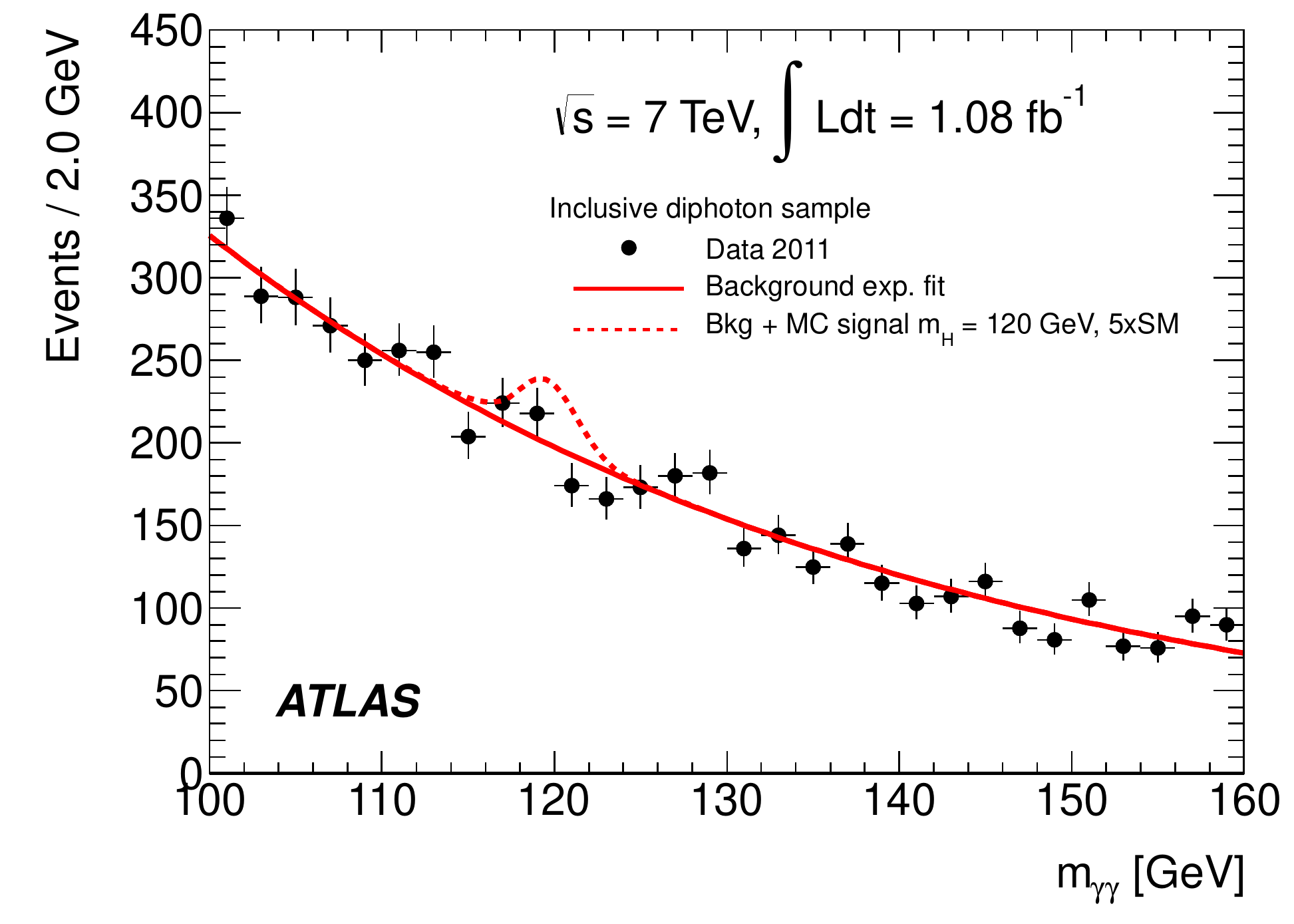}
\includegraphics[width=0.49\textwidth]{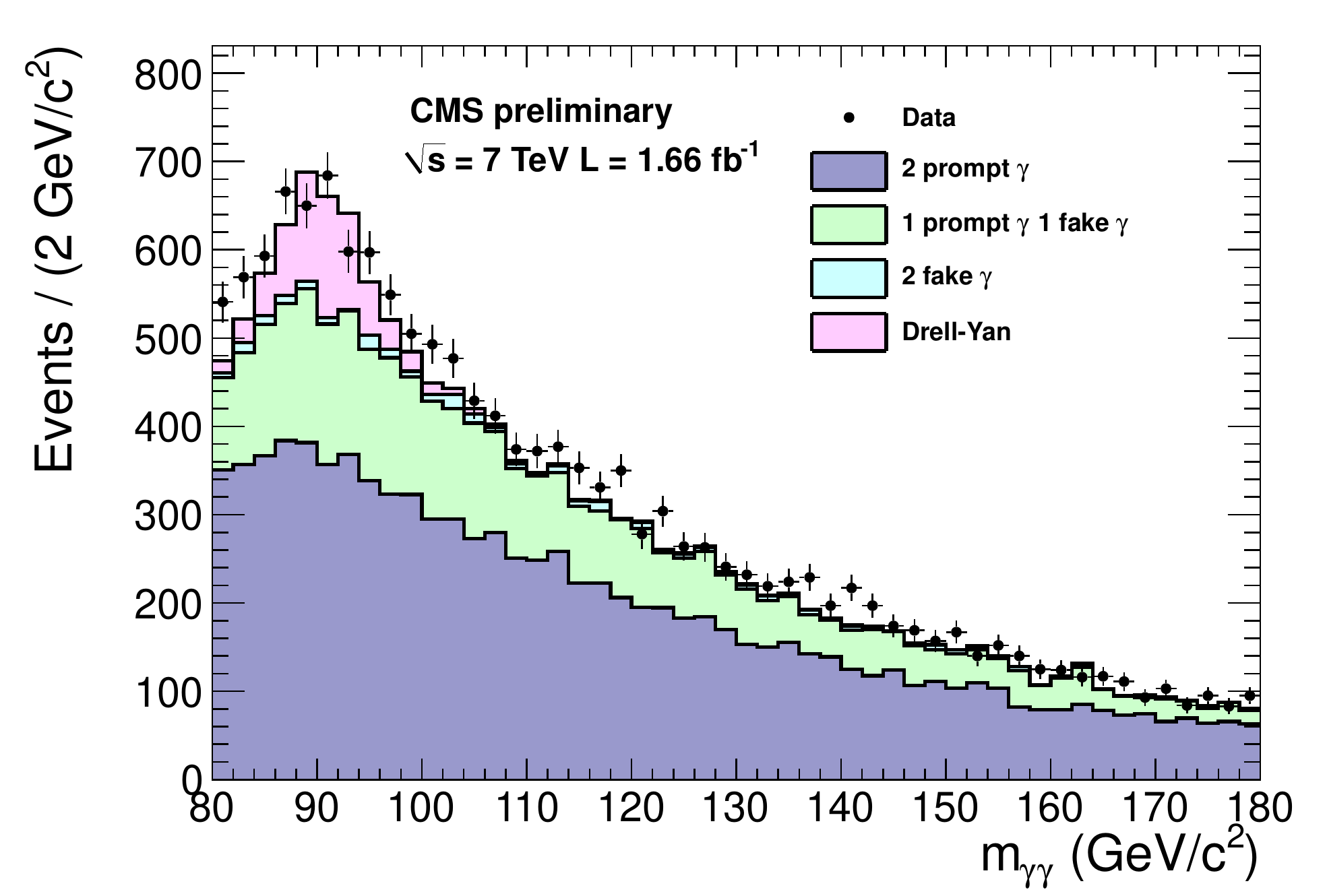}
    \vskip -0.2cm
    \caption[]{The two-photon mass spectrum as measured by ATLAS
      (left) and CMS (right).  }
    \label{fig:hgg}
  \end{figure}

The background modelling is done using exponential functions (ATLAS)
or Bernstein polynomials (CMS). Each experiment has sensitivity to
three to four times the SM 
 signal rate, remarkably consistent given the different design
 choices.
 There are no deviations of more than two sigma from the
 background expectations at present.

\subsection{Higgs boson decay to WW}

The largest decay mode of the SM Higgs boson is to W boson pairs,
which approaches 100\% for masses around 165~\GeV. The purely- and
semi-leptonic decay modes have both been searched for using the leptons
as triggers. 

\subsubsection{$H \rightarrow $ \WWlnqq}

The largest branching ratio is in \WWlnqq, which ATLAS
has reported on in Ref.~\cite{:2011as}. The
properties of the single missing neutrino are estimated using
missing energy constraints and requiring the reconstructed W to be on
mass shell, which results in a quadratic equation  for \pz. 
This allows the mass to be calculated and the
search becomes a counting experiment like $H\rightarrow \gamma
\gamma$, looking for a peak on a smooth background. 

  \begin{figure}[h]
\centering
\includegraphics[width=0.49\textwidth]{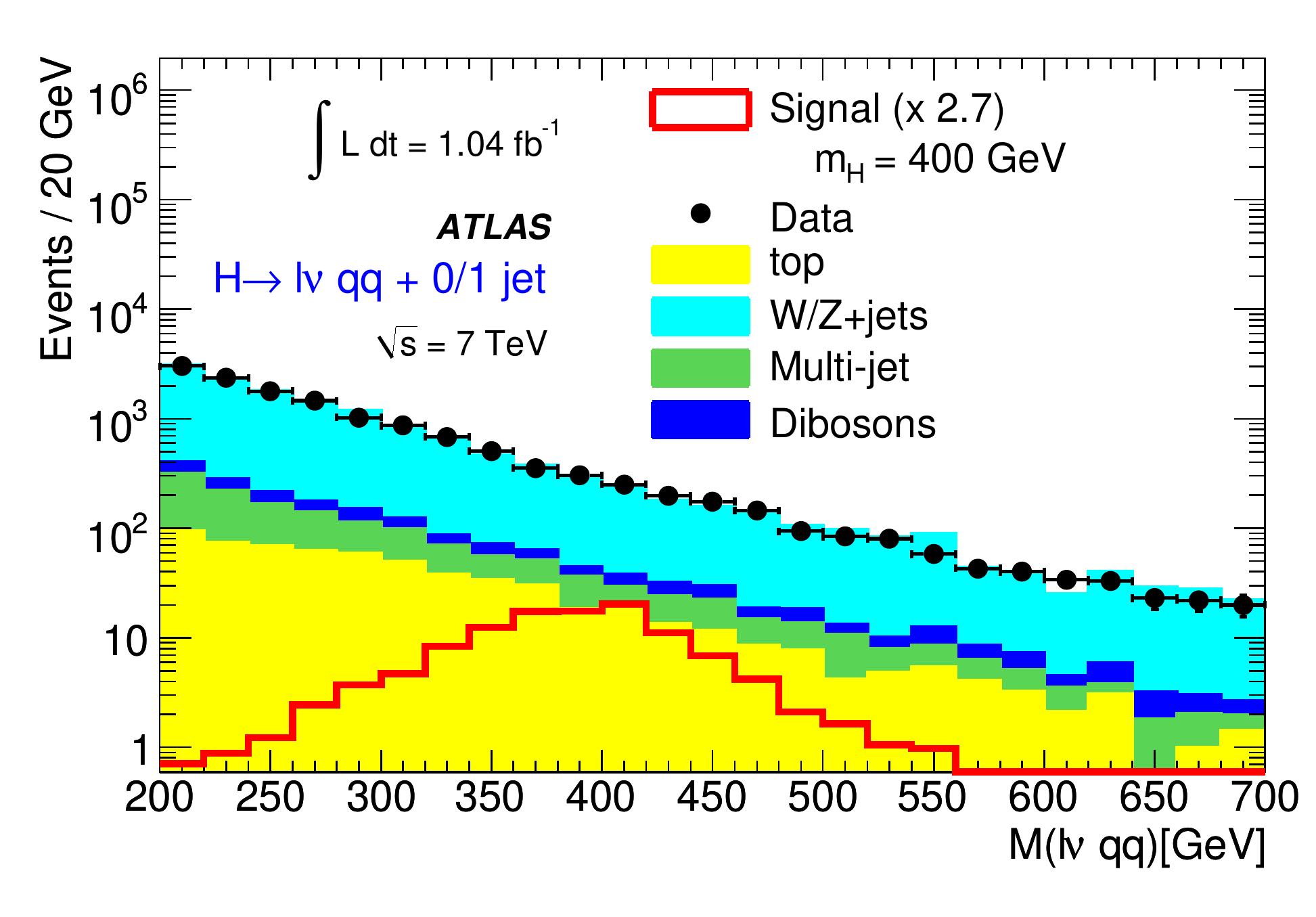}
\includegraphics[width=0.49\textwidth]{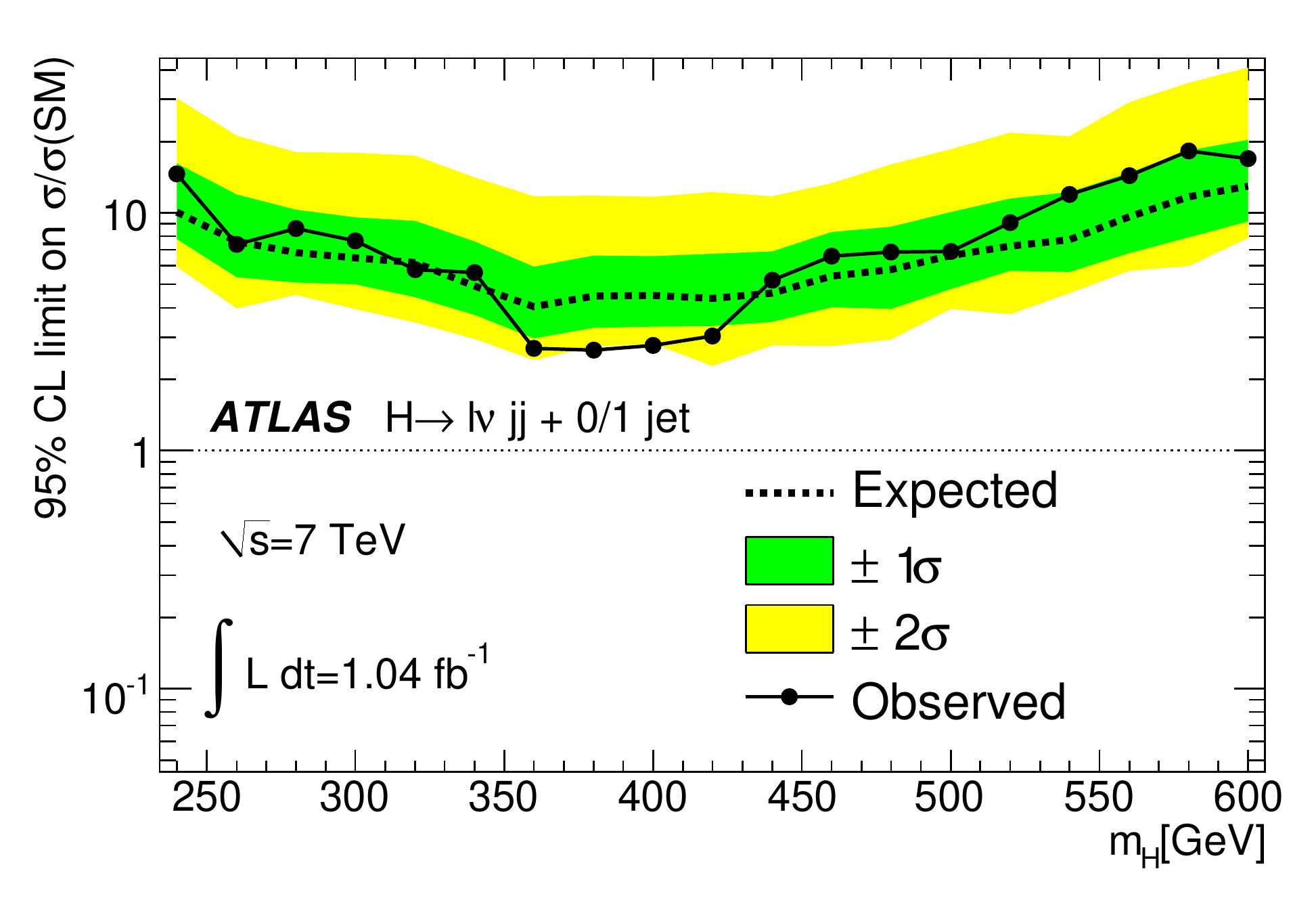}
    \vskip -0.2cm
    \caption[]{Left: The observed mass spectrum in the ATLAS
      $H\rightarrow WW \rightarrow \l\nu qq$ search. Right limits on
      the Higgs
boson       productions rate, in units of the SM Higgs
boson rate,
      coming from this search.}
    \label{fig:hWWlnqq}
  \end{figure}

The results, shown in Fig.~\ref{fig:hWWlnqq} show sensitivity to a
cross-section about 2.7 times that of the SM. The problem for this
search is the very large W plus jets background  which masks any
potential signal

\subsubsection{$H \rightarrow $\WWlnln}

The most powerful search at present is the doubly leptonic
WW decay. The presence of two leptons and substantial missing energy
allows for excellent QCD suppression, and the spin-zero nature of the Higgs
boson aligns the spin of the  Ws and hence  the decay leptons tend to be
emitted in similar directions, while the dominant WW background does
not have this feature\cite{Dittmar:1996ss}.

The search is done by first removing, so far as possible, backgrounds
other than $WW$ production via tight identification of two isolated
leptons and missing energy.  Lepton pairs compatible with the $Z$
mass are excluded  and the events are divided into subcategories
dependent upon the number of associated jets. In the  0 jets category
the background is 
dominantly non-resonant WW, while the 1 jet events have a sizeable top
contribution
and in the exclusive two 
jet case, used only by CMS,  additional cuts on the rapidities of the
jets are applied to enhance sensitivity
to the vector boson fusion production process. As a final step the
ATLAS analysis\cite{wwNote} selects a region in the transverse
mass\cite{Barr:2009mx}, while the CMS collaboration, in addition, uses a
boosted decision tree to select candidates\cite{CMS-hww}.

  \begin{figure}[h]
\centering
\includegraphics[width=0.49\textwidth]{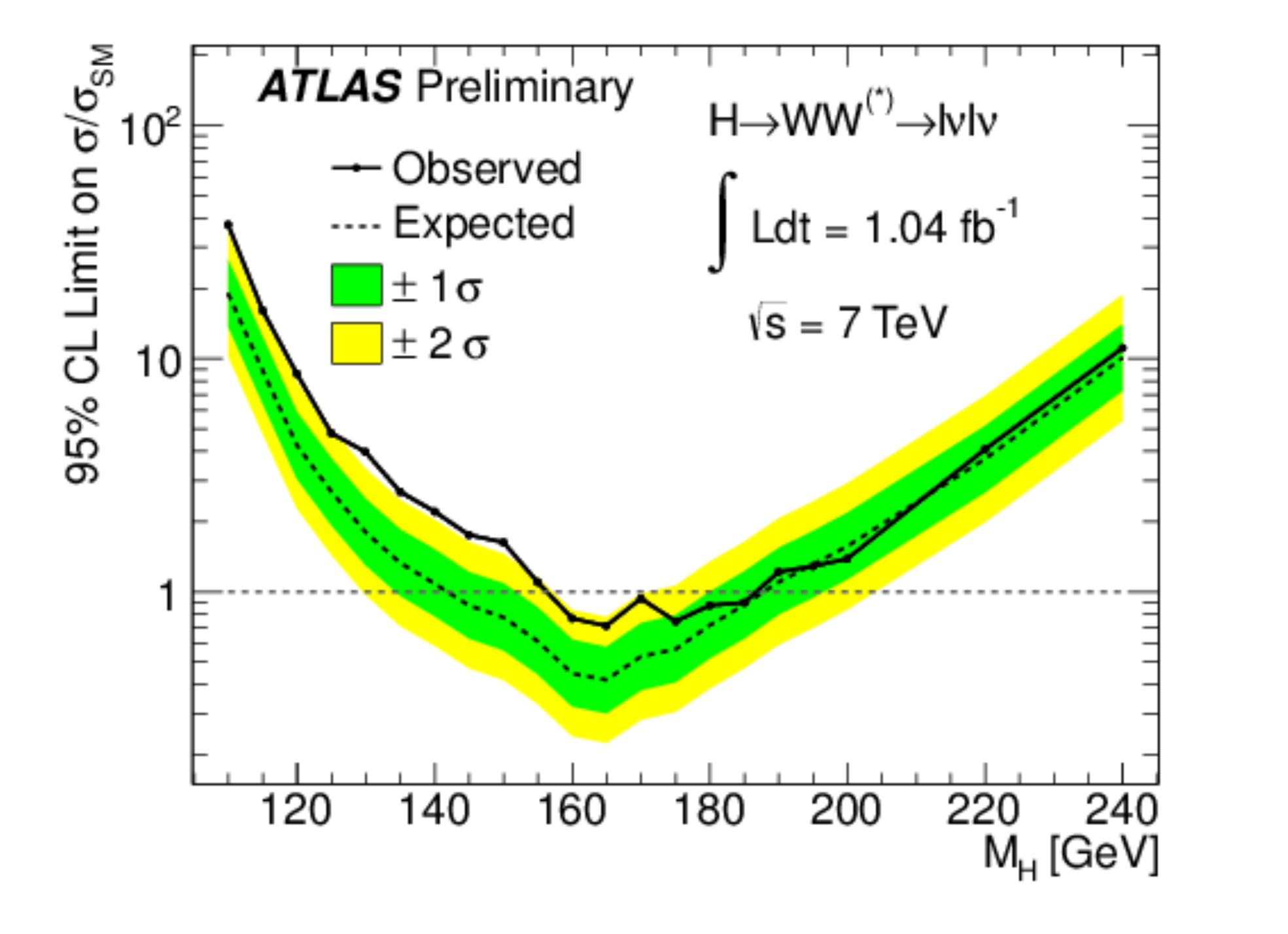}
\includegraphics[width=0.49\textwidth]{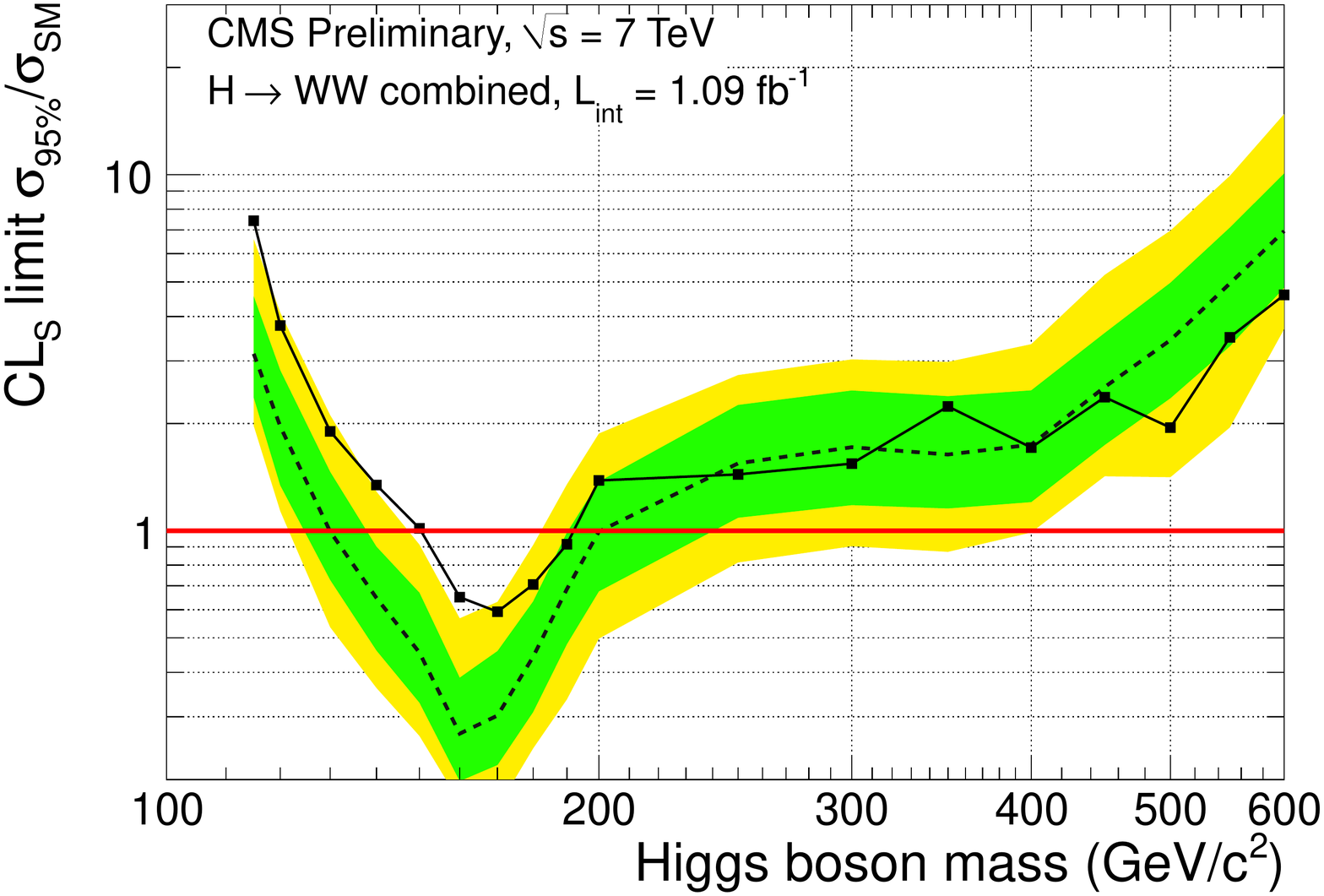}
    \vskip -0.2cm
    \caption[]{The multiple of the SM rate which can be
      excluded by ATLAS (left) and CMS(right) in the $h\rightarrow WW
      \rightarrow l\nu l\nu$ search as a function of \mh.}
    \label{fig:hwwlvlv}
  \end{figure}

This channel has important  sensitivity, with expected exclusions of
142~\GeV\ to 186~\GeV\ in ATLAS and 130~\GeV\ to 200~\GeV\ in CMS. The
observed exclusions are rather less, being 158-186~\GeV\ in ATLAS and
150-193~\GeV\ in CMS. The reason for the discrepancy between expected
and observed is a noticeable excess of candidates, in both
experiments. The mass resolution is very poor, owing to the two
missing neutrinos, and this excess manifests itself over a wide mass region.

\subsection{Higgs boson decay to ZZ}

The decay to pairs of Z bosons has the  potential for very clean
searches due to the 
attractive features of the subsequent Z boson decays:
particle-antiparticle pairs allow cross-checks and it is often
possible to use the well-known Z mass as a powerful  constraint.
ATLAS and CMS have both studied events where one Z decays to leptons
and the other to leptons, neutrinos or quarks\cite{epsZZatlas,epsZZcms}.

\subsubsection{$H \rightarrow$\ZZllll}

The cleanest channel for the Higgs boson search at the LHC is the decay to pairs
of Z bosons with their subsequent decay to electron or muon pairs\cite{epsZZatlas,epsZZcms,llllNote,cms-llll}. The
multiple constraints of four clean leptons and one or two resonant Z
bosons means that the selected sample is dominated by real diboson
production, with potentially a narrow Higgs boson signal in addition.
 When looking for Higgs bosons weighing less than twice the
Z mass then one of the Z bosons must be off mass shell, and the
kinematic selection is looser; this can be partially compensated by
tightening the lepton identification criteria, but in general the
emphasis is on maximising the signal efficiency while preserving the
low background.

The background is dominated by the irreducible  non-resonant ZZ
production, for which the 
shape is   predicted by MCFM\cite{mcfm6}. There are
complications when one boson is significantly off mass shell and for
production from gluon induced processes; both of these affect the low
mass region especially. This region also has significant
backgrounds from Z plus jets and \ttbar. These are constrained from
the data. The observed mass spectra,
and predicted backgrounds, can be seen in Fig~\ref{fig:llllcands}.

  \begin{figure}[h]
\centering
\includegraphics[width=0.49\textwidth]{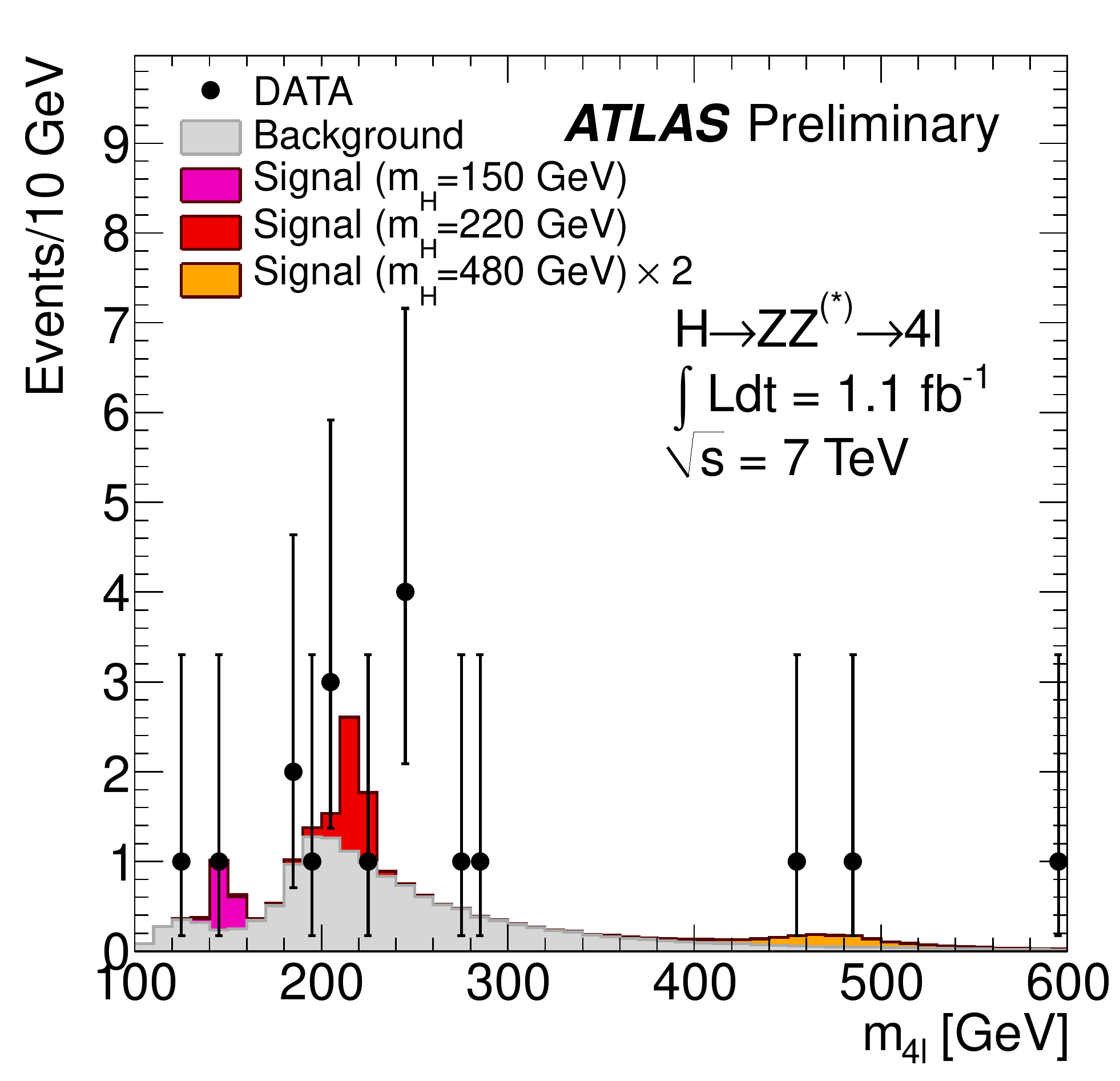}
\includegraphics[width=0.49\textwidth]{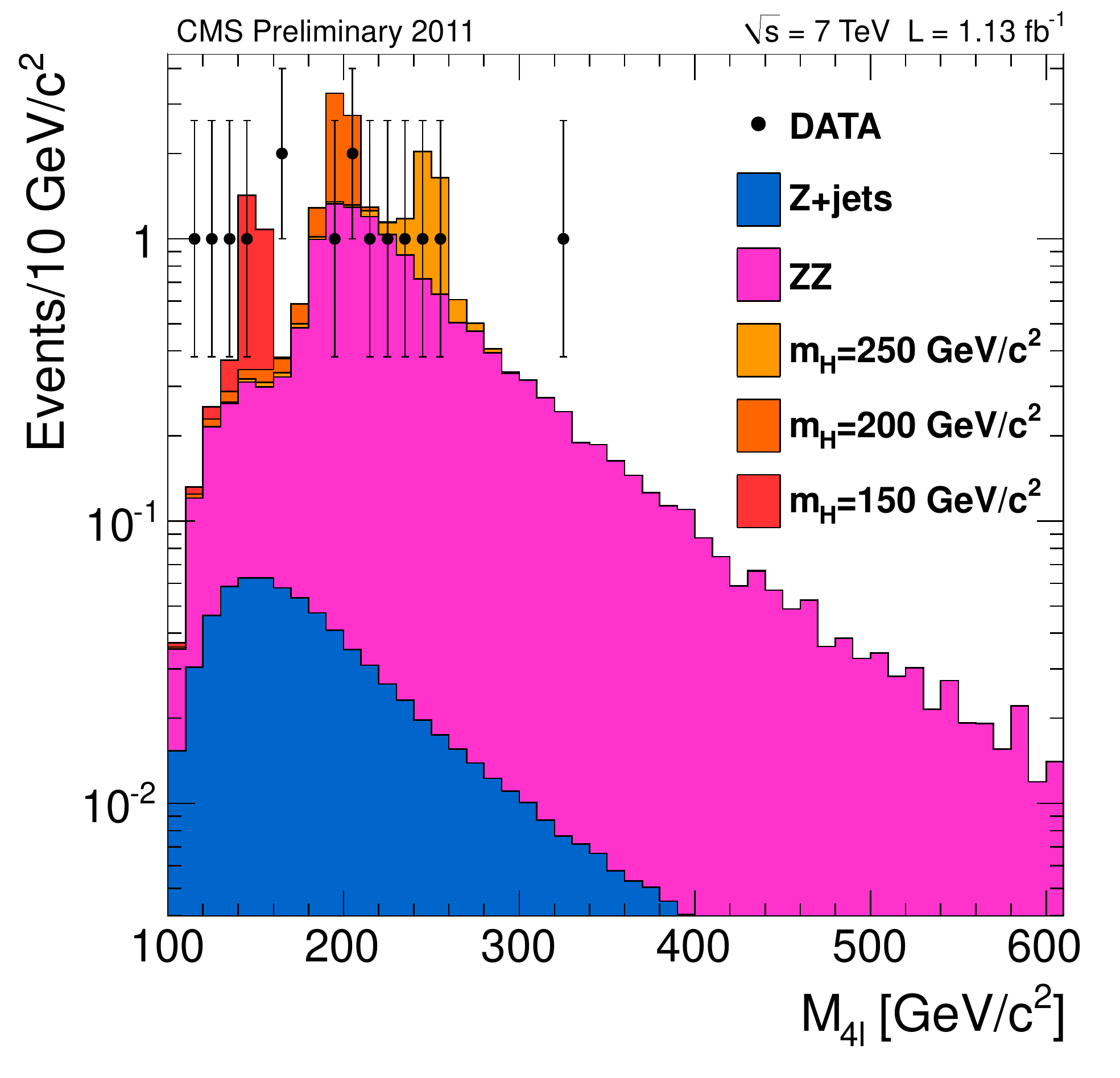}
    \vskip -0.2cm
    \caption[]{The selected event candidates in the $H \rightarrow$
    \ZZllll\  search of ATLAS (left) and CMS (right). The
      CMS curve breaks the background into ZZ and Z plus jets; 
      similar fractions are seen in ATLAS.}
    \label{fig:llllcands}
  \end{figure}

The sensitivity of these searches is close to being able to exclude
the SM Higgs boson 
at 200~\GeV, but with the current luminosity a very small mass region
is excluded by this result alone. 
Attention is drawn by the low mass events; ATLAS and CMS taken together have 
 three candidates near 143~\GeV which is consistent with the expectation
for a signal there.

\subsubsection{$H \rightarrow $\ZZllqq}

The $llqq$ decay mode has a higher rate than a purely leptonic decay,
but also allows complete reconstruction of the Higgs boson
candidates. Indeed it has much in common with the decay \WWlnqq.
 One Z boson is reconstructed through its decay to
electrons or muons and the other is looked for as a pair of jets.
The ZZ rate (from a Higgs boson or non-resonant) is so much smaller than the
Z plus jets background that it is not  possible to observe the peak
of the Z in the jet jet continuum. Thus a window is selected which
should, in simulation, contain most of the second Z events. One effect
of this is that if the jet energy scale is incorrectly measured in
either direction the efficiency for selecting signal candidates
reduces.

Having obtained two candidate Z bosons, the putative Higgs boson mass is
easily reconstructed, and the distribution of these masses is tested
for a signal. The sensitivity is improved by tagging b-quarks as they
are produced in a much larger fraction of ZZ decays than in the Z plus
jets background. The experiments use b-tagged and untagged candidates
separately. CMS further subdivides to have a category of events with
one clear b-tag and also rejects events from the untagged sample where
the jets are identified by a gluon jet tagger. They also employ a likelihood
discriminant based upon the angles of the decay to reject backgrounds\cite{cms-llqq}.
ATLAS uses different kinematic selections depending
upon whether the Higgs boson mass hypothesis tested is above or below
300~\GeV\cite{Aad:2011ec}.
The mass distributions in the (high-mass) b-tagged channel can be
seen in Fig~\ref{fig:llqq}, which provides approximately half of the total
sensitivity. 

  \begin{figure}[h]
\centering
\includegraphics[width=0.53\textwidth]{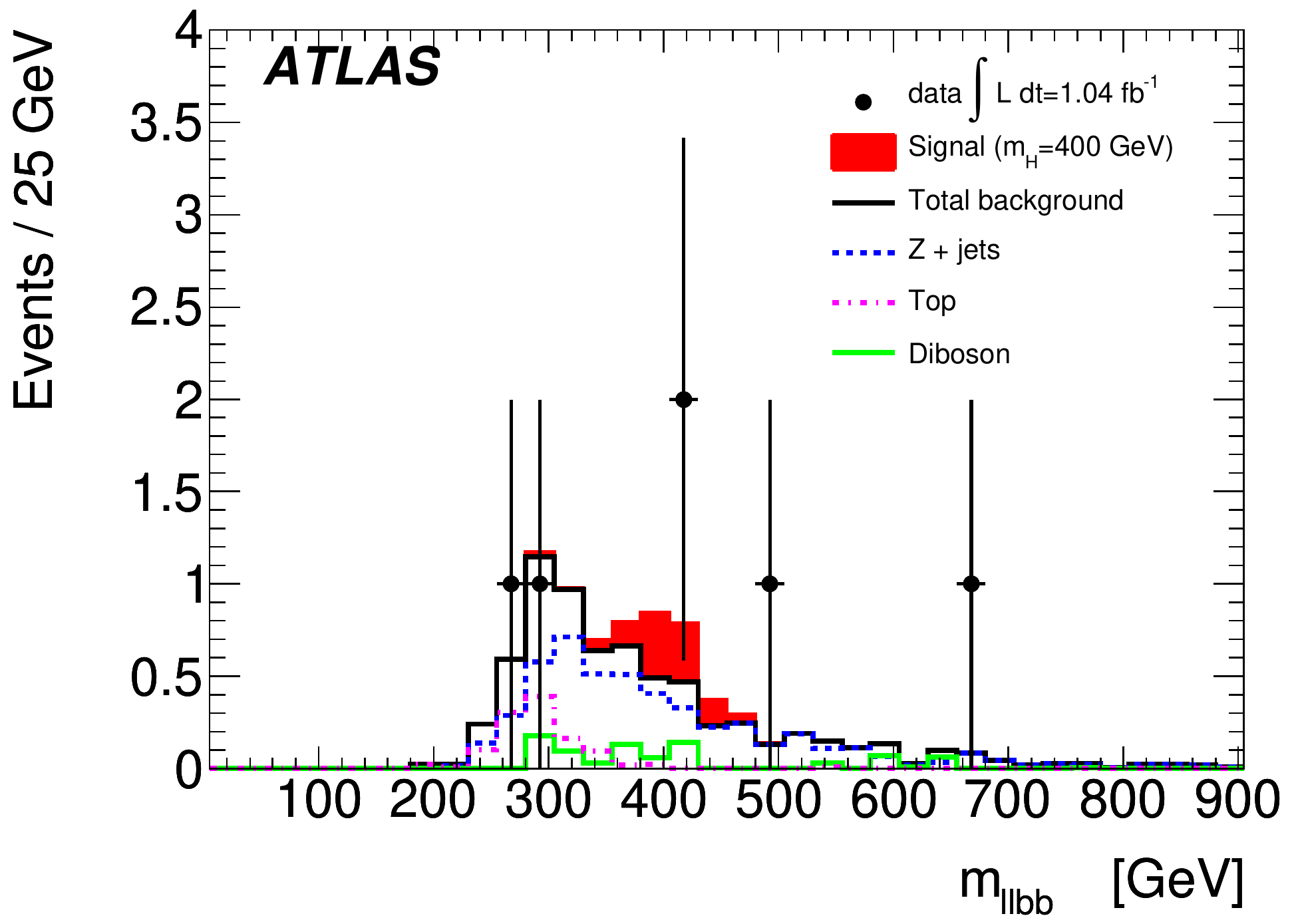}
\includegraphics[width=0.45\textwidth]{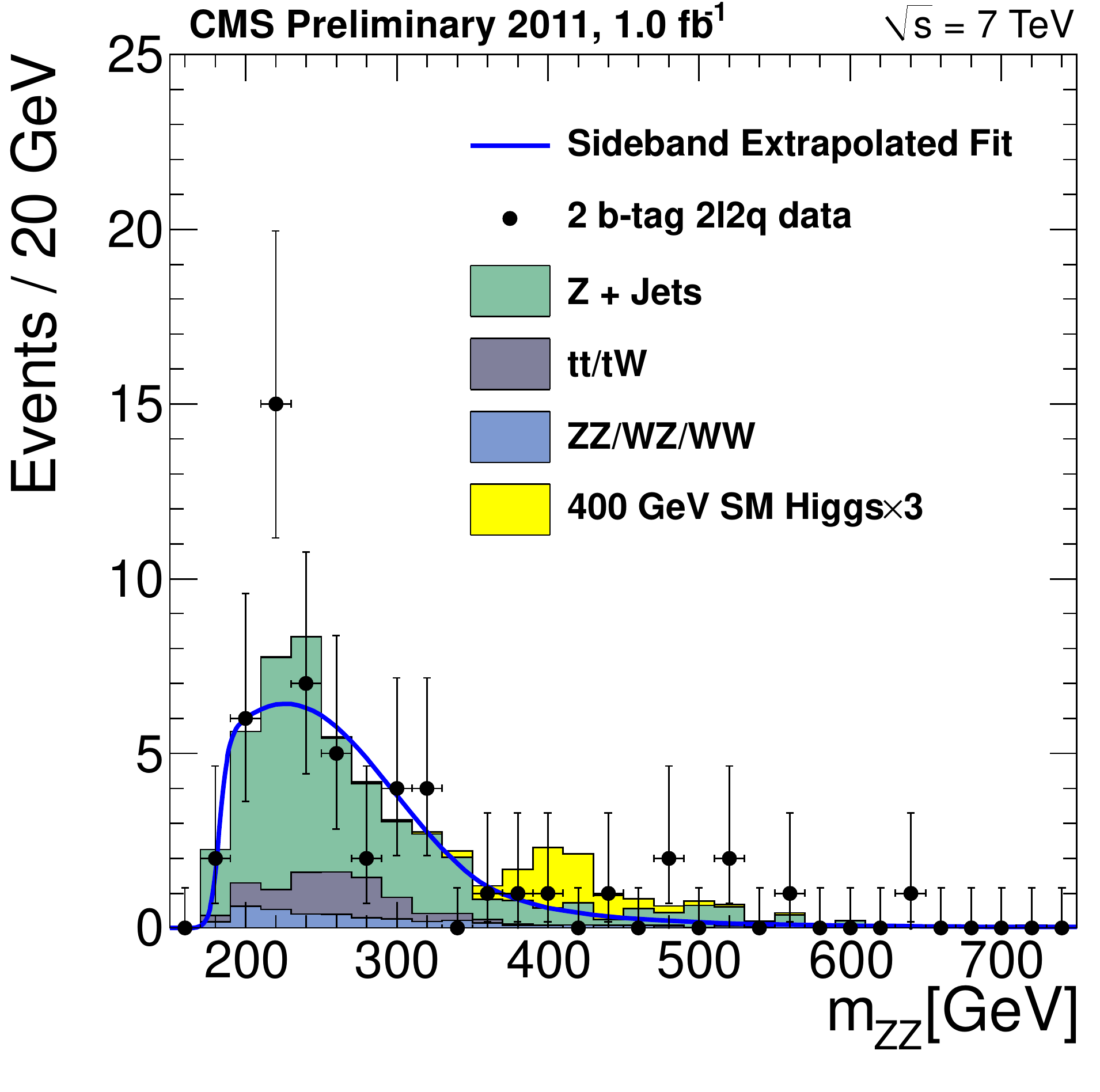}
    \vskip -0.2cm
    \caption[]{The mass distributions measured by  ATLAS (left) and
      CMS (right) in the search for \ZZllqq. In each case only the
     $\ell\ell  b b$ distribution is shown here. } 
    \label{fig:llqq}
  \end{figure}

The \ZZllqq\ searches do not reveal any striking excess, never
departing from the two-sigma expected region.  Limits are set which
are at best about 1.7 times the \SM\ cross-section.

\subsubsection{$H \rightarrow $\ZZllnn}

The ZZ Higgs boson decay mode, with subsequent decay of one Z to
charged leptons and the other to neutrinos has many attractive
features for an LHC search. The two hard leptons  give a good trigger
signature, and the purely leptonic signature allows excellent
background rejection. however, unless the \pT\ of the Z which decays
to neutrinos is large it is hard to detect, and the background from
inclusive Z production is overwhelming. For large Higgs boson masses
this is easily satisfied and this becomes the most sensitive analysis
for Higgs boson masses over about 300~\GeV. 

The reconstruction of the Higgs boson mass is  not possible in the
presence of two neutrinos, but by assuming they arise from the decay
of an on-shell Z boson the transverse mass can be reconstructed. An
important item then is the modelling of backgrounds from single Z
boson production, where ATLAS and CMS have taken different
strategies. ATLAS models this in their simulation, and makes some
verification's on the data that the performance is as
expected\cite{:2011va}, while 
CMS uses $\gamma$ events to measure the missing \ET\ distribution and
correct for the difference in mass to obtain the expected missing
\ET\ distribution in Z events\cite{cms-llnn}. The transverse mass distributions
measured by the two experiments can be seen in Fig.~\ref{fig:llnn}.

  \begin{figure}[h]
\centering
\includegraphics[width=0.49\textwidth]{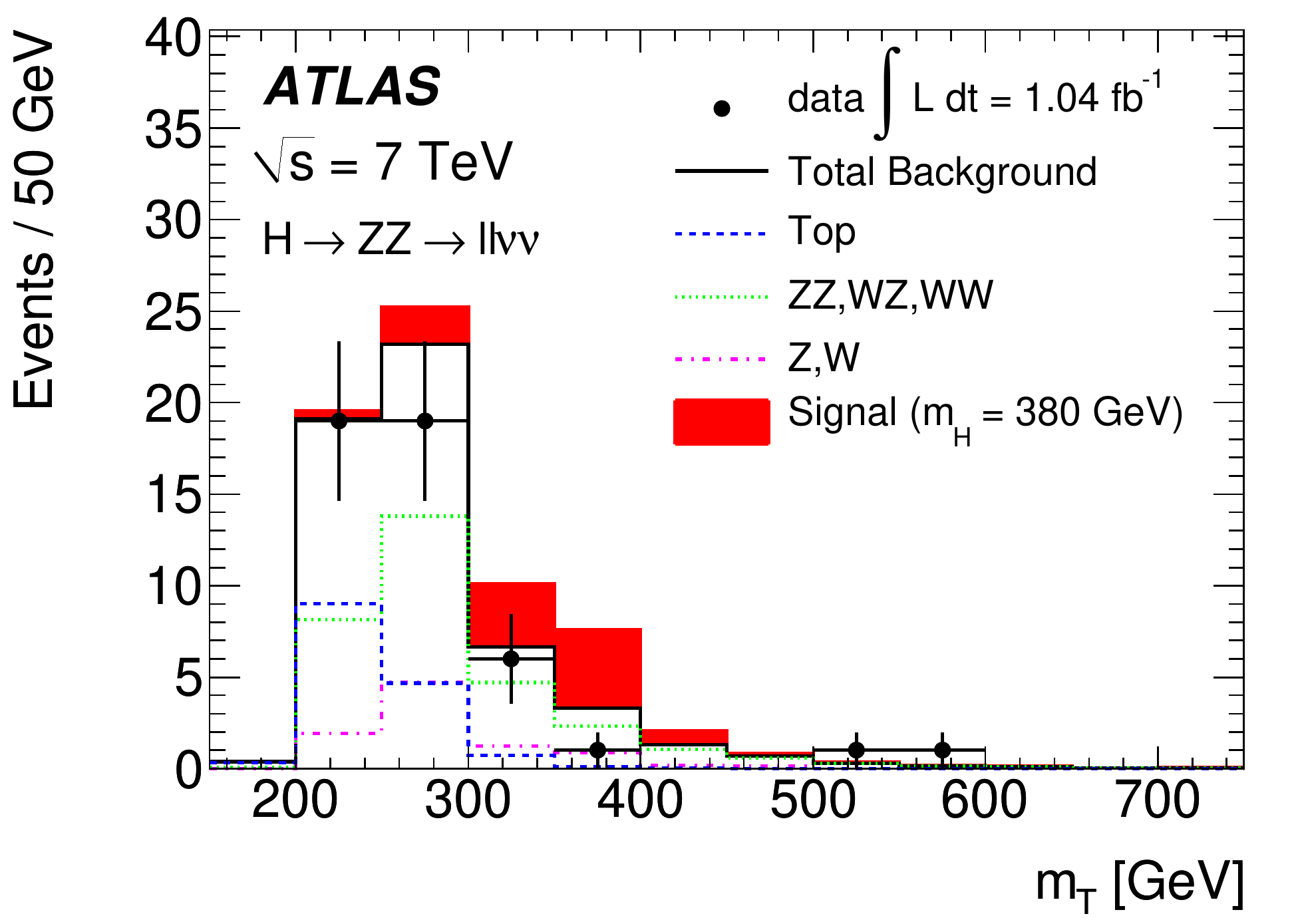}
\includegraphics[width=0.49\textwidth]{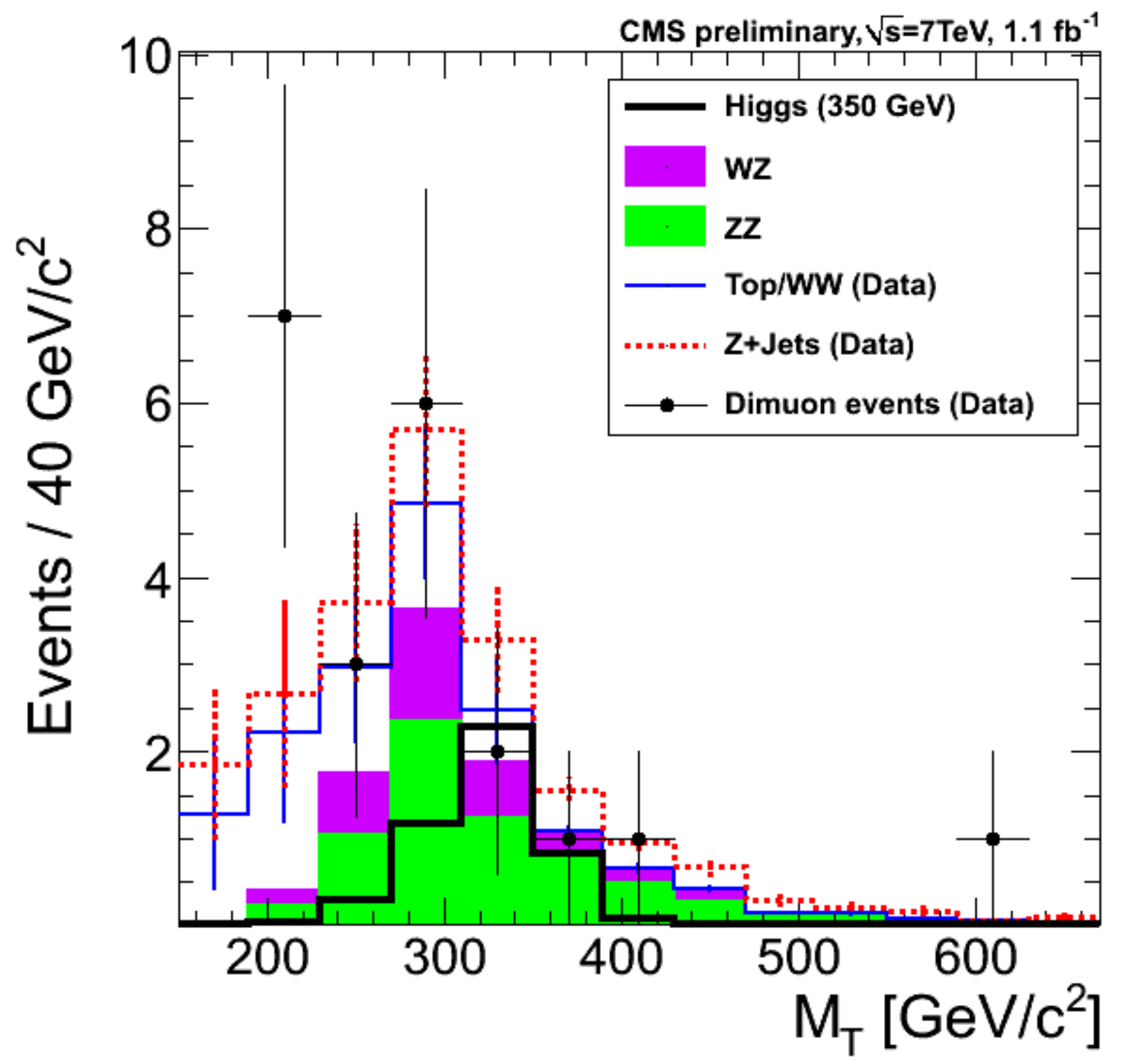}
    \vskip -0.2cm
    \caption[]{The mass distributions observed by ATLAS (left) and CMS
      (right) in the $h\rightarrow$
      \ZZllnn\  search. The ATLAS distribution includes both lepton
      flavours, in the case of CMS, only the muon channel is shown.}
    \label{fig:llnn}
  \end{figure}

\subsection{Combined search results}

The only unknown parameter in the SM Higgs boson search is its mass;
this makes it ideally suited to a combination approach where all the
disparate channels are considered together. 
Each channel has its own region of applicability, as shown in
Table~\ref{ta:channels}.  
The results of the
individual channels in the two experiments are displayed in
Fig~\ref{fig:channels}. 
The most powerful channels are common to the two experiments, but at
low mass ATLAS has analysed the \bbbar\ decay mode while CMS searches
for the $\tau\tau$ decay. At high masses CMS has used the
\WWlnln\ decay mode while ATLAS looks at \WWlnqq. 

\begin{table}[h]
\begin{center}
\begin{tabular}{lcccc}
\hline
  & \multicolumn{2}{c}{ATLAS} & \multicolumn{2}{c}{CMS} \\
  & Luminosity & Mass range  & Luminosity & Mass range \\
  & \ifb\     &  ~\GeV         &    \ifb\  & ~\GeV  \\
\hline
 $\gamma\gamma$              & 1.08 & 110-150 & 1.1 & 110-140 \\
 $\tau\tau$                  & -    &  -      & 1.1 & 110-140 \\
 \bbbar                      & 1.04 & 110-130 & -   & - \\
\WWlnln                      & 1.04 & 110-240 & 1.1 & 110-600 \\
\WWlnqq                      & 1.04 & 240-600 & -   & - \\
 \ZZllll                     & ~1.1 & 110-600 & 1.1 & 110-600 \\
\ZZllnn                      & 1.04 & 200-600 & 1.1 & 250-600 \\
\ZZllqq                      & 1.04 & 200-600 & 1.0 & 226-600 \\
\hline
\end{tabular}
\end{center}
\caption{The channels used by each of the experiments, along with the
  amount of data  contributing to them  and the mass range for which they produce results.}
\label{ta:channels}
\end{table}

  \begin{figure}[h]
\centering
\includegraphics[width=0.49\textwidth]{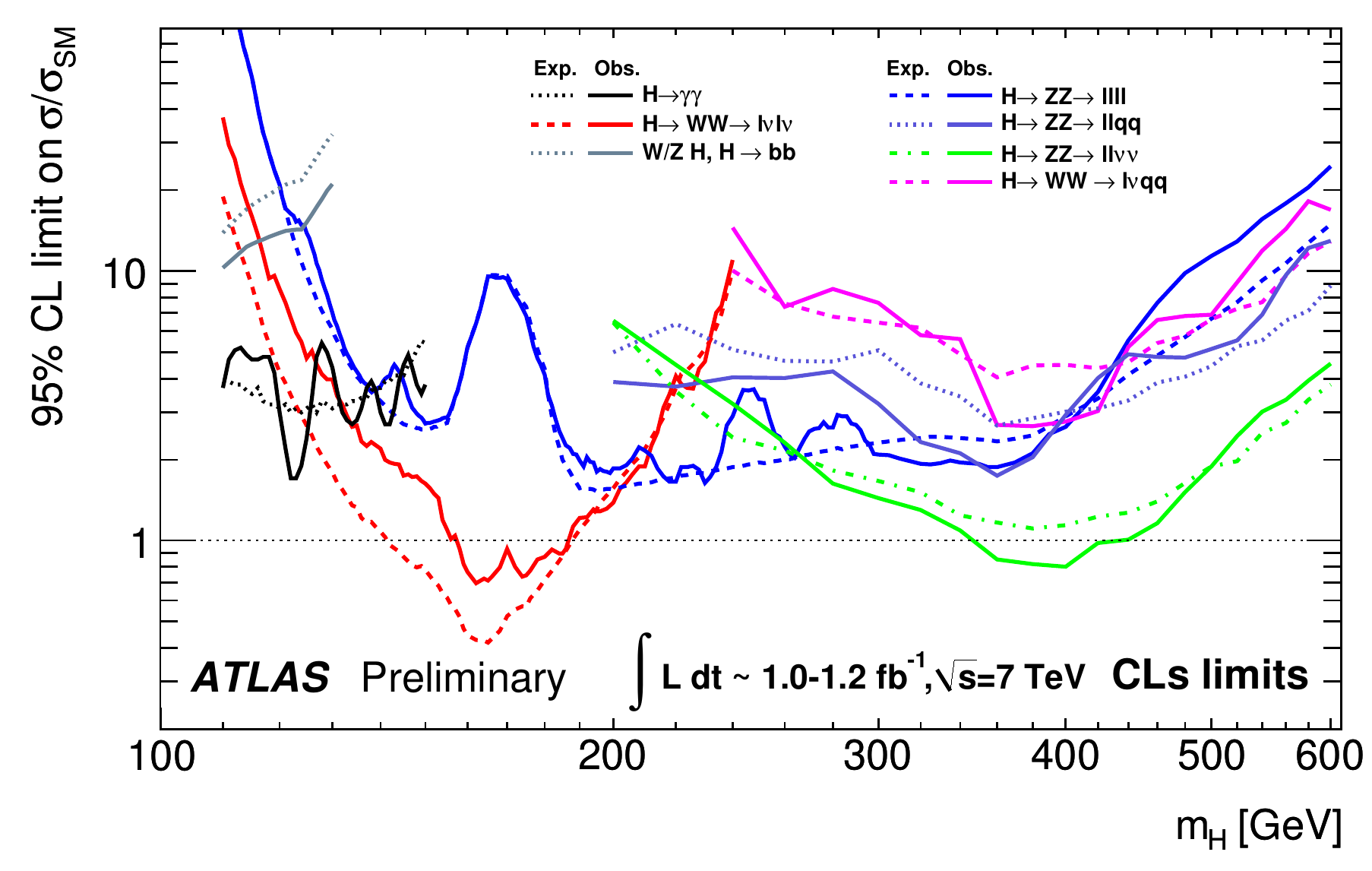}
\includegraphics[width=0.49\textwidth]{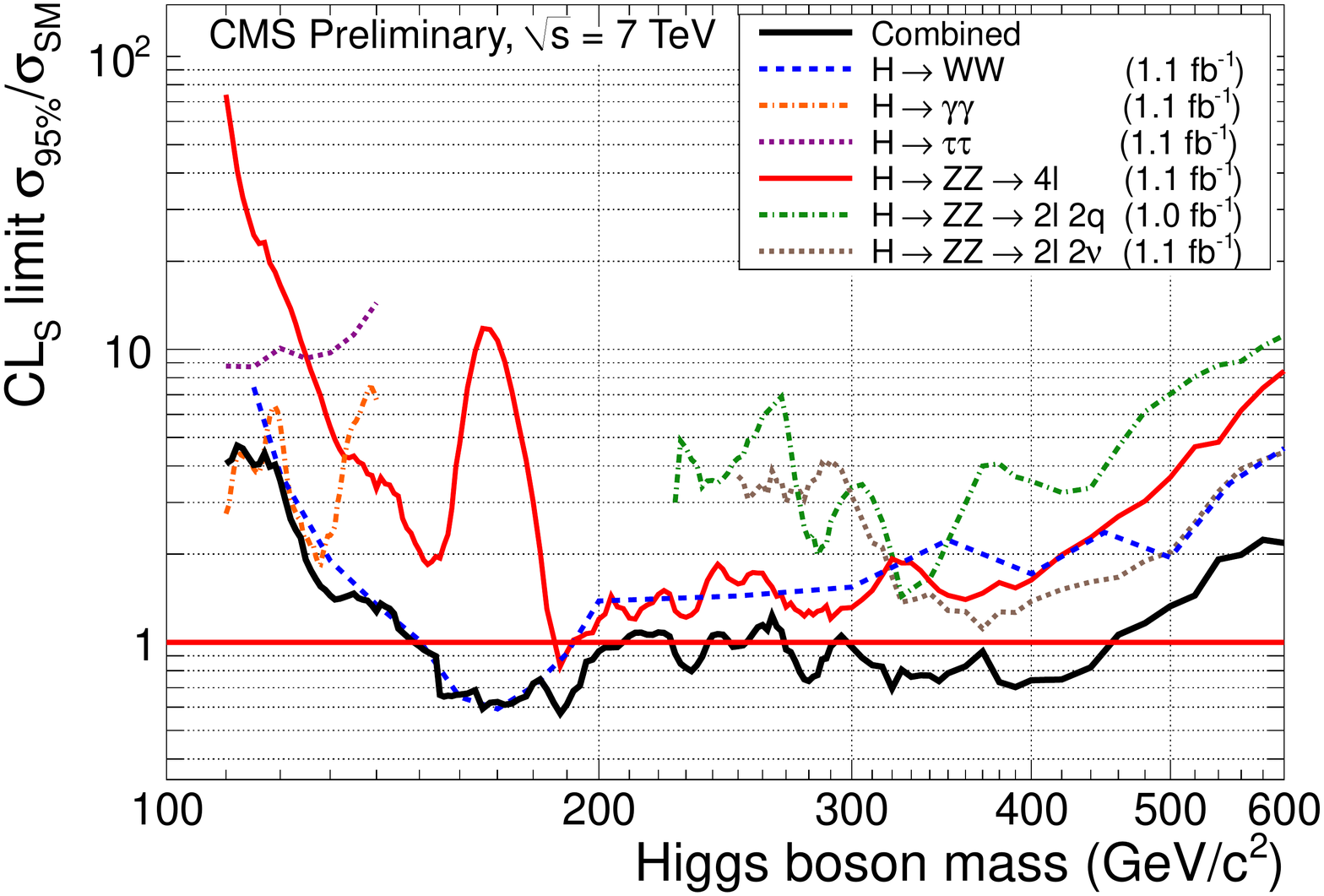}
    \vskip -0.2cm
    \caption[]{The multiple of the SM rate which is
      excluded by ATLAS (left) and CMS(right) by the individual
      Standard Model search channels. ATLAS show the expected as well
      as the observed limits, while  CMS display the combined
      result, which excludes  a large region at high mass where no
      single channel does. } 
    \label{fig:channels}
  \end{figure}

In many respects the two experiments are very similar. The
sensitivity is largely from the $\gamma \gamma$ search for Higgs boson
masses below about 120~\GeV, but then the \WWlnln\ search dominates to
200~\GeV with \ZZllll\ then taking a major role. 
For Higgs boson masses above about 300~\GeV\ the \ZZllnn\ search has
the greatest 
power. One striking fact is that the \WWlnln\ search by CMS at
high mass is important to very high masses and is not even considered
by ATLAS.

Both collaborations have produced  combinations of their own results,
and the limits extracted from these searches can be seen in
Fig.~\ref{fig:comb}.

  \begin{figure}[h]
\centering
\includegraphics[width=0.49\textwidth]{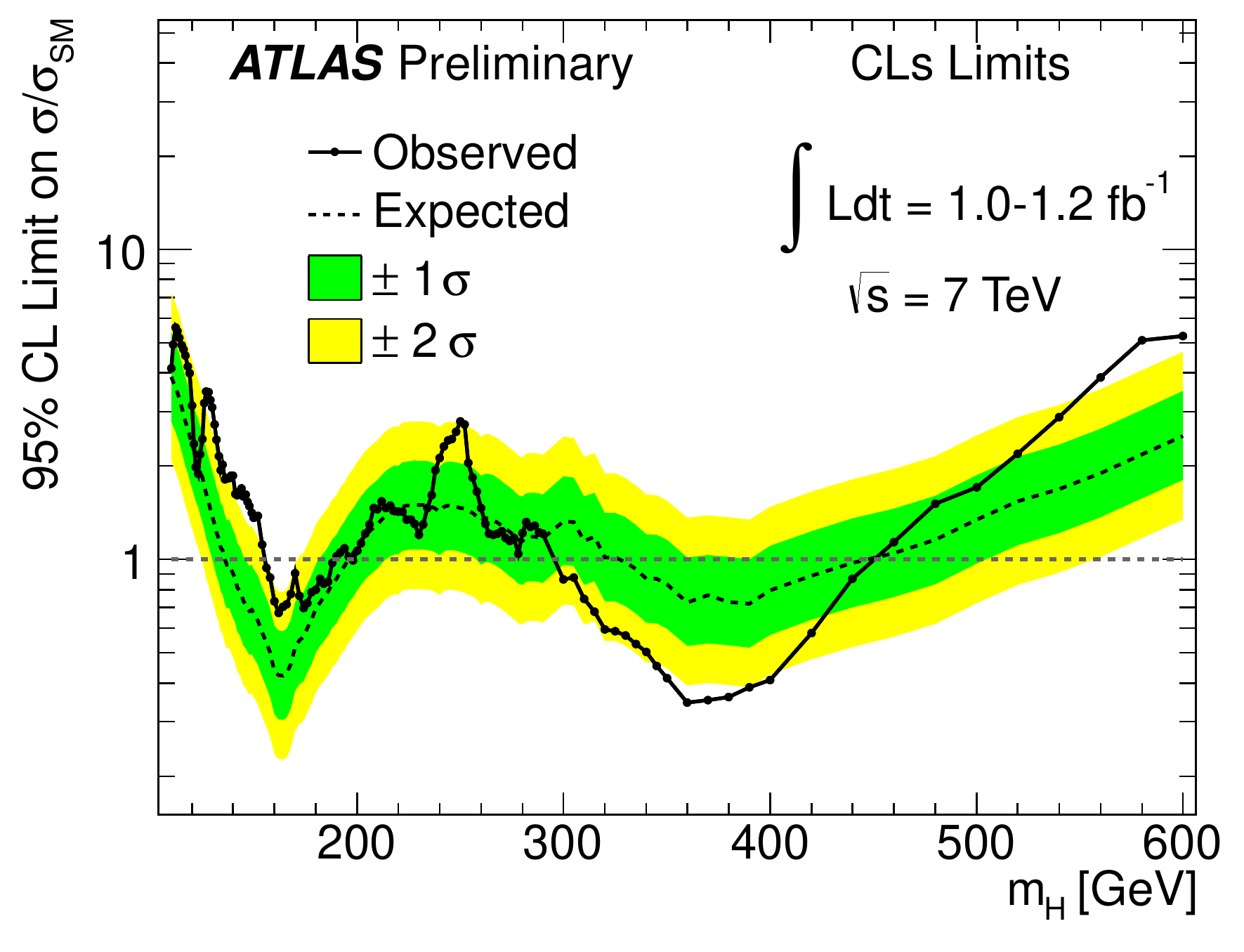}
\includegraphics[width=0.49\textwidth]{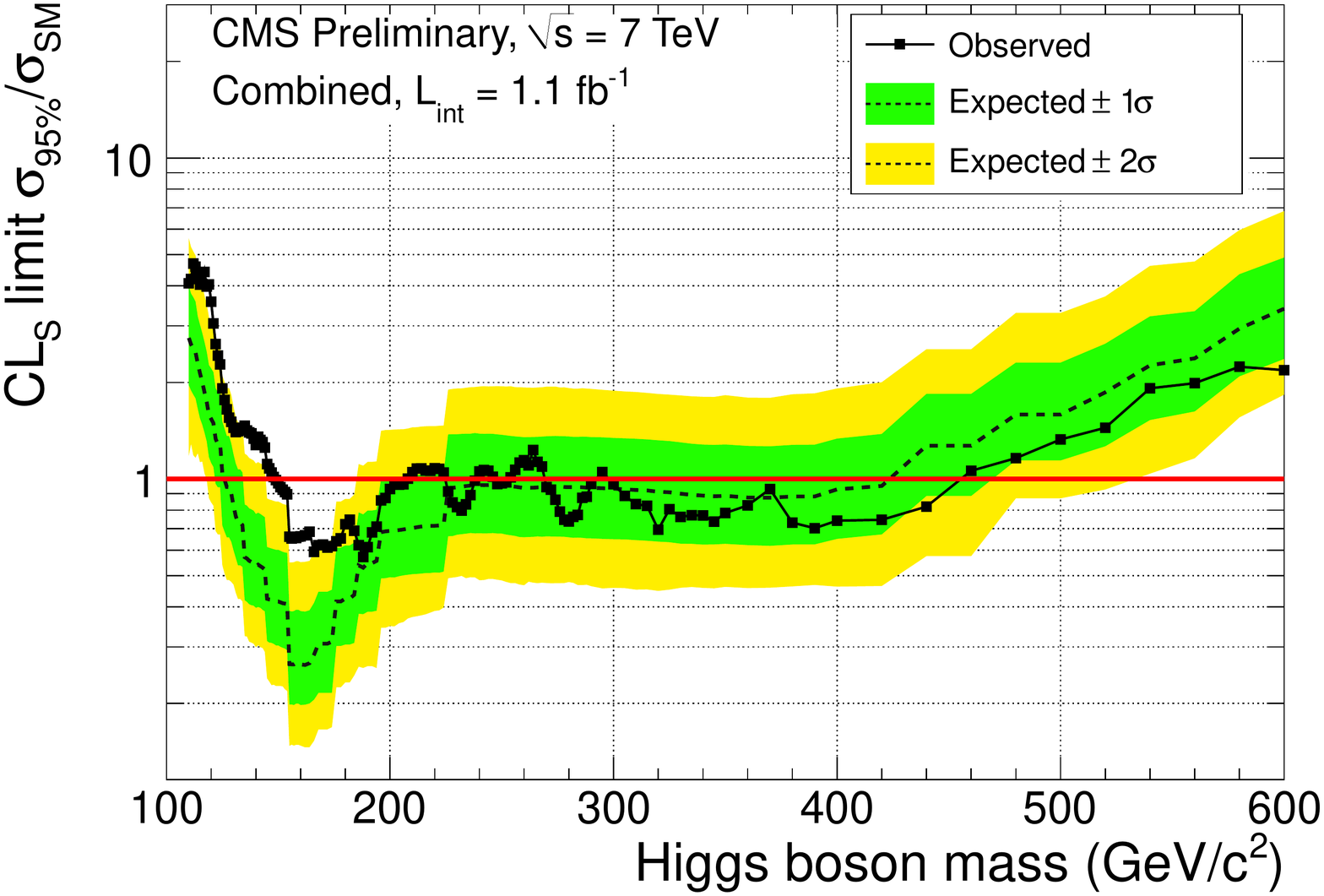}
    \vskip -0.2cm
    \caption[]{The multiple of the SM rate which can be
      excluded by ATLAS (left) and CMS(right) by the combination of
      Standard Model search channels used.}
    \label{fig:comb}
  \end{figure}

 The  sensitivity is
high across a wide range of Higgs boson masses, and ATLAS excludes the
mass regions 155 to 190~\GeV\ and 295 to 450~\GeV, while CMS rules out
149 to 206~\GeV, 270 to 290~\GeV\ and 300 to 400~\GeV, all at 95\% CL.

The expected sensitivity of the two experiments can be compared. The
low mass region, dominated by $\gamma\gamma$, both are very similar, while
the CMS sensitivity around 160~\GeV\ is noticeably better. This comes
at least in part from the use of a multivariate analysis by CMS in the
\WWlnln\ search while ATLAS use a cut-based approach. Around
250~\GeV\ the CMS search is again more powerful largely due to the same WW
search which was optimised also for this region, but also from a more
sensitive \ZZllll\ search. Above 340 \GeV\ ATLAS
has greater sensitivity, largely coming from the \ZZllnn\ channel.

The other striking feature of the data is the difference between the
observed and expected limits at low mass. This is occasioned by an
excess of candidates compared with expectations, and is a feature of
the data of both ATLAS and CMS, and can be seen in Fig~\ref{fig:pval}.
At high masses ATLAS has excesses at 246~\GeV\ and 580~\GeV; neither
of these are strongly seen in the CMS data.

  \begin{figure}[h]
\centering
\includegraphics[width=0.49\textwidth]{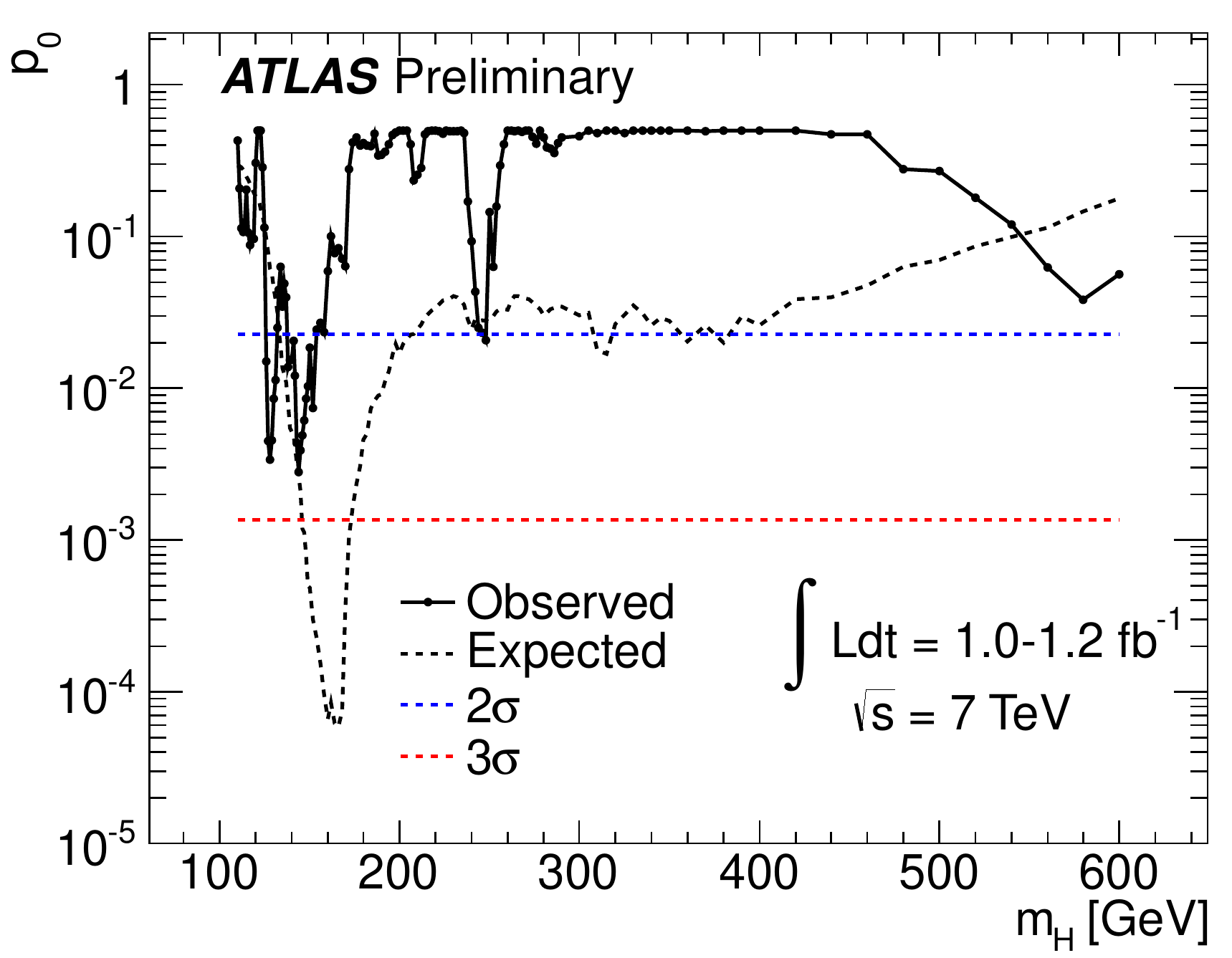}
\includegraphics[width=0.49\textwidth]{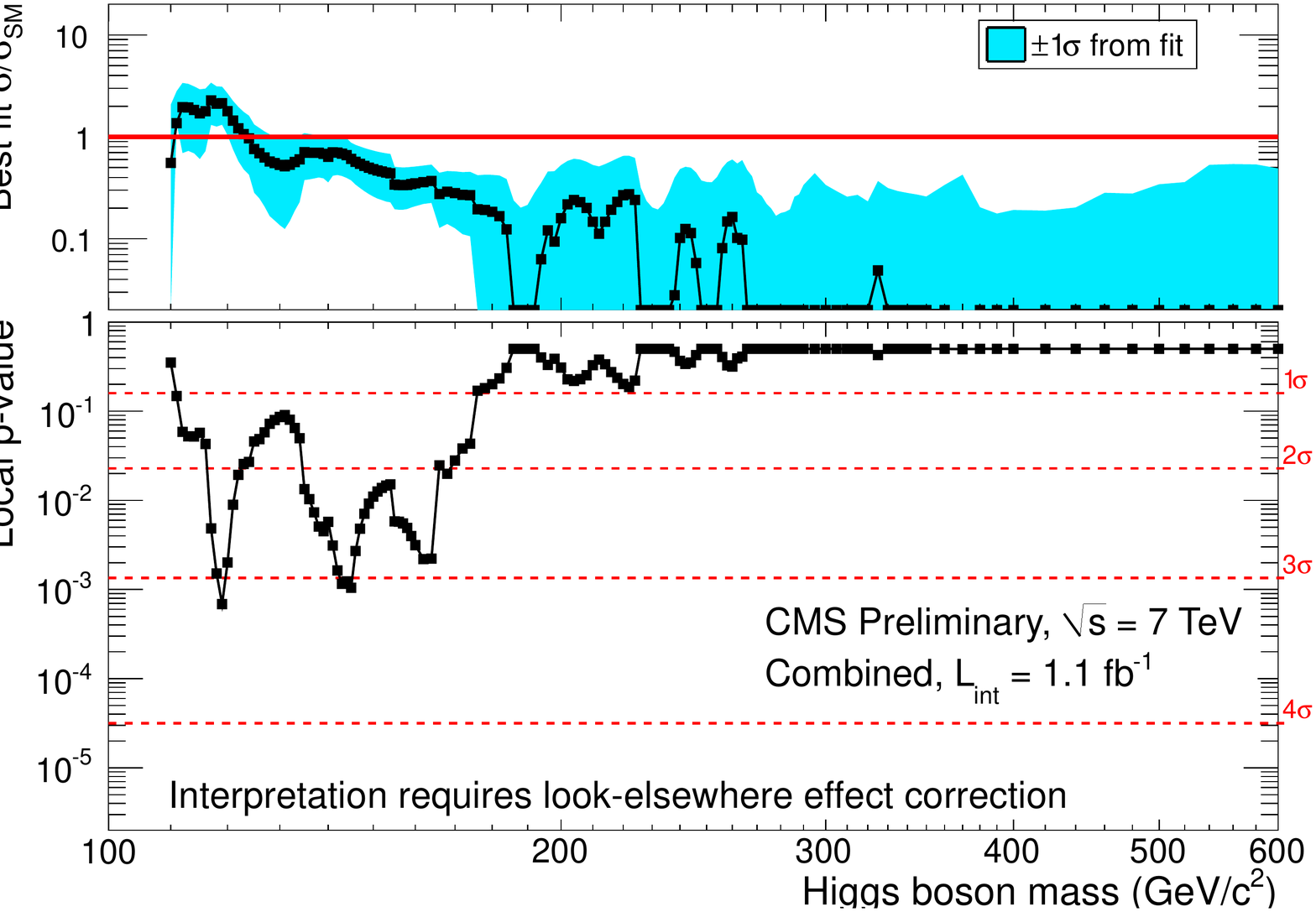}
    \vskip -0.2cm
    \caption[]{The probability of getting as signal-like an excess in
      the presence of only backgrounds, as a function of the Higgs
      boson mass, for ATLAS (left) and CMS (right). Deficits appear at 0.5 by
      construction.  ATLAS also show the expected evidence if there
      were a signal, while CMS show the strength  of the observed
      excess in units of the \SM\ Higgs boson cross-section. No
      look-elsewhere effect is considered in these plots. }
    \label{fig:pval}
  \end{figure}

The low mass situation, see in Fig~\ref{fig:pval2}, reflects a broad
excess in both channels. This is mostly driven by 
the $WW$ searches, but is given definition by
peaks in the $\gamma\gamma$ mass spectrum and individual candidates in
the \ZZllll\ search. ATLAS has local minima at 128 and 145~\GeV\ while
CMS has similar  at 119, 146 and 164~\GeV. The match between the peaks near
145~\GeV\ is interesting, but more data will be required to disentangle
the situation.

  \begin{figure}[h]
\centering
\includegraphics[width=0.6\textwidth]{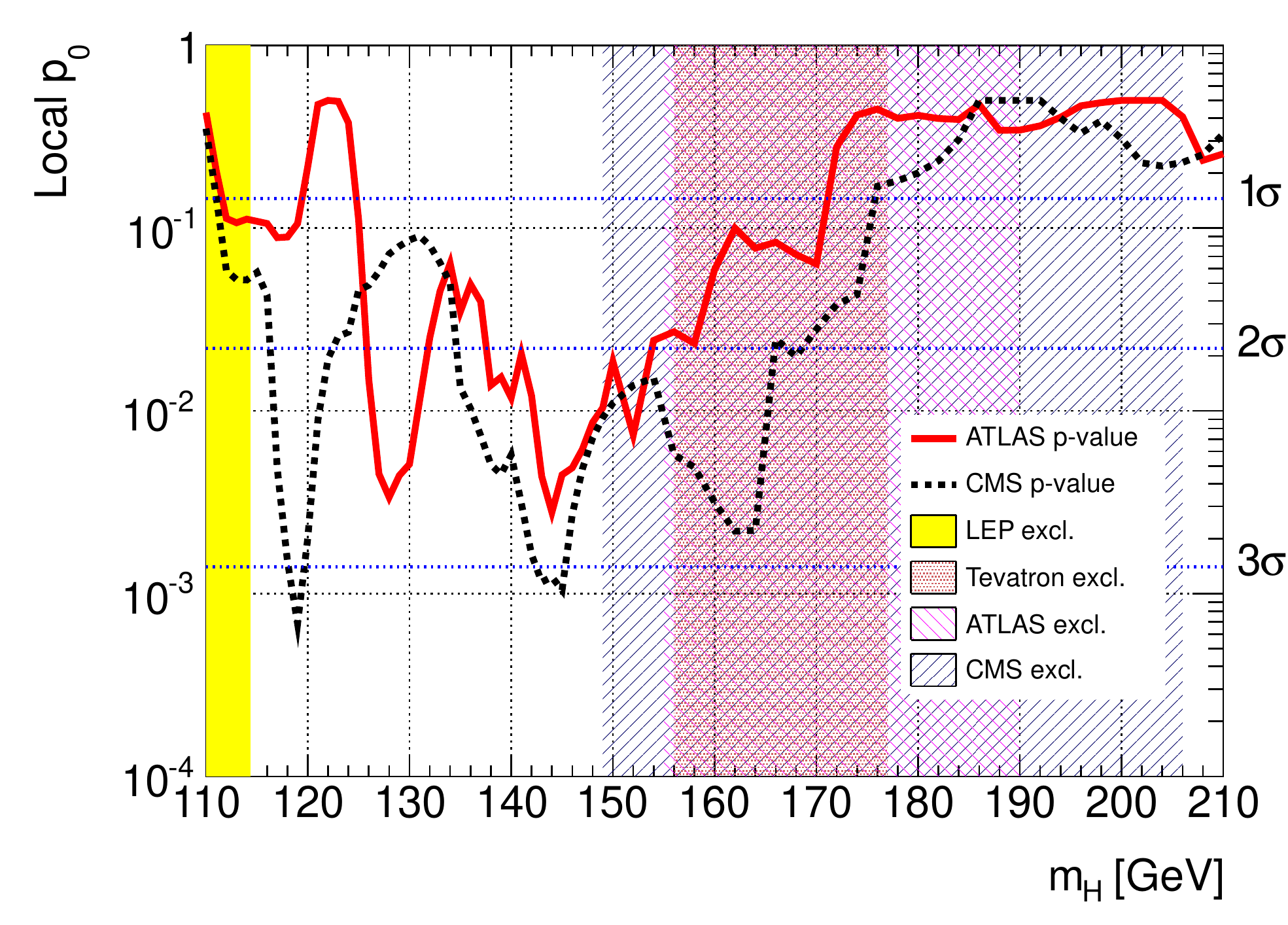}
    \vskip -0.2cm
    \caption[]{The probability of getting as signal-like an excess in
      the presence of only backgrounds, as a function of the Higgs
      boson mass, for the two experiments. Deficits appear at 0.5 by
      construction. No look-elsewhere effect is considered in this plot.}
    \label{fig:pval2}
  \end{figure}

\section{Outlook}

The first inverse femtobarn of integrated luminosity from the LHC have
produced a huge sensitivity to the SM Higgs boson, with median
exclusion sensitivities of both experiments at, or close to, 95\% CL for
all masses between 135~\GeV\ and 450~\GeV. Most of this region is in fact
excluded, but the upper limits on a low mass Higgs boson set by  ATLAS
and CMS, 155~\GeV\ and 149~\GeV\ are noticeable higher than might be
expected, reflecting mostly the excess of events observed in the $H
\rightarrow  $WW searches. These are of course the most sensitive
search channel for much of this range, but whether this is the start of
an observation, systematic bias, or merely random fluctuation will
best be resolved by more data.

The LHC performance continues to be excellent and it is clear that
substantially more data will be collected in 2011 and beyond which
will answer the question of the existence or otherwise of the
SM Higgs boson. 
 
\bibliographystyle{unsrt}
\bibliography{eps_wjm}
%

\end{document}